\documentclass[reprint,
 amsmath,amssymb,
 aps, superscriptaddress, pra
]{revtex4-2}

\usepackage{amsmath}
\usepackage{graphicx}
\usepackage{dcolumn}
\usepackage{bm}
\usepackage[usenames,dvipsnames]{xcolor}
\usepackage{cases}
\usepackage{float}
\usepackage[caption = false]{subfig}
\usepackage[normalem]{ulem}

\newcommand{\microspace}{\mspace{0.5mu}}
\def\<{\langle}
\def\>{\rangle}
\def \lket {\left|}
\def \rket {\right\rangle}

\def \IPR {\mathcal{I}}
\def \IPRavg {\overline\IPR}
\def \NumMeas {\mathcal{N}}
\newcommand{\ket}[1]{\lket\microspace #1 \microspace\rket}

\def \nt {\tilde n}

\renewcommand{\vec}{\mathbf}

\begin{document}
\title{Measurement resolution enhanced coherence for lattice fermions}

\author{Hilary~M.~Hurst}
\affiliation{Department of Physics and Astronomy, San Jos\'{e} State University, San Jos\'{e}, California, 95192, USA}
\author{Yik~Haw~Teoh}
\affiliation{Department of Physics and Astronomy, San Jos\'{e} State University, San Jos\'{e}, California, 95192, USA}
\author{I.~B.~Spielman}
\affiliation{Joint Quantum Institute, National Institute of Standards and Technology, and University of Maryland, Gaithersburg, Maryland, 20899, USA}

\date{\today}

\begin{abstract}
Weak measurement enables the extraction of targeted information from a quantum system while minimizing decoherence due to measurement backaction.  However, in many-body quantum systems backaction can have unexpected effects on wavefunction collapse.  We theoretically study a minimal many-particle model consisting of weakly measured non-interacting fermions in a one dimensional lattice.  Repeated measurement of on-site occupation number with single-site resolution stochastically drives the system toward a Fock state, regardless of the initial state.  This need not be the case for measurements that do not, even in principle, have single-site spatial resolution.
We numerically show for systems with up to 16 sites that decreasing the spatial resolution strongly affects both the rate of stochastic evolution for each quantum trajectory and the allowed final states. The full Hilbert space can be partitioned into backaction-free subspaces (BFSs) the elements of which are indistinguishable to these measurements.  Repeated measurements will drive any initial state into a single BFS, leading to a steady state that is a fixed point of the measurement process.
We exactly calculate the properties of these BFSs for systems up to 32 sites and find that even for moderate reductions in measurement resolution they yield non-trivial steady state entanglement and coherence.
\end{abstract}

\maketitle

\section{Introduction \label{Sec:Intro}} 

The interaction of a quantum system with its environment generally disturbs its evolution, leading to an unavoidable loss of information.  
Repeated quantum measurements act as a special type of environment where the observer retains a measurement record, and thereby obtains information about individual quantum systems over time. These `measurement-environments' offer intriguing possibilities for exploring new phenomena in quantum systems. For example, when combined with feedback, measurement can support reservoir engineering by steering a system into some specific desired state~\cite{Diehl2008}. 
Even without feedback, the recently identified measurement-induced entanglement phase transitions (EPTs) illustrate the potential for creating non-trivial (but not targeted) quantum states by combining entangling operations with decoherence processes in the form of partial quantum measurements~\cite{skinner2019measurement, Gullans2020, lavasani2021measurement,ippoliti2021entanglement}. In loose analogy with measurement based quantum computation~\cite{Briegel2009}, here we focus on the yet more simplified case of an initially entangled state of spinless fermions subject to repeated weak and partial measurement of the spatial density distribution. Even with no additional dynamics, we find that the final state conditioned on the measurement record can retain significant coherence. This minimal model is ideally suited for implementation using an ultracold atomic Fermi gas in a 1D optical lattice: the initial entangled state is no more than the Fermi sea in a half-filled lattice. Weak global measurements can be implemented using established techniques~\cite{andrews1997propagation, Ketterle1999, shin2006observation, shin2007tomographic, Altuntas2023}; and these measurements are made partial by the natural resolution limit governed by the optical wavelength.

Fermions provide a very unusual setting in which to explore measurement-based quantum control.  
Typically these ideas arise in a quantum information context, and are most developed for quantum circuits~\cite{Gullans2020, lavasani2021measurement, ippoliti2021entanglement, iadecola2023measurement, Agrawal2022} and spin systems~\cite{Ma2018, Garcia2019, munoz2020simulation, lang2020entanglement}.
A handful of results discuss EPTs for quantum gases~\cite{Cao2019,Chen2020emergent,buchhold2021effective,alberton2021entanglement}, but those proposals require the usual combination of measurement and entangling dynamics.
Other quantum control protocols developed for ultracold atoms have focused on feedback cooling and many-body state preparation~\cite{hurst2020feedback, Young2021, Yamaguchi2022, Wu2022, Wampler2022, mehdi2023fundamental}. 
Fermionic systems remain relatively unexplored, in part because measurement is complicated by Fermi statistics: a measurement cannot drive a fermion into a state which is already filled~\cite{pezze2004insulating, kohl2005fermionic, moritz2005confinement, giorgini2008theory}.

We explore the question of residual spatial coherence in a minimal model of immobile, non-interacting fermions on a 1D lattice undergoing weak quantum measurements of atomic density.
We show that the effect of the optical resolution limit on an initial Fermi sea is evident both in the time to achieve steady state and nature of that state.
For higher resolutions the steady states are simple Fock states with no spatial entanglement, but at a critical resolution limit non-trivial superpositions begin to appear.  In this way, the measurement resolution can tune between regimes with remnant coherence and those without, indicating a type of coherence transition.
We uncover how the measurement protocol partitions the full Hilbert space into smaller backaction-free subspaces (BFSs), which can allow the steady states to preserve remnant coherence and entanglement. These results bridge quantum information theory and experimental implementations of measurement-based control, providing a way forward for conditional quantum state engineering and control in Fermi systems. 

The paper is organized as follows: In Sec.~\ref{Sec:Model} we augment a model of weak homodyne measurements~\cite{hurst2019measurement} to include finite spatial resolution, and introduce our approach for numerically exact simulations of this measurement process. In Sec.~\ref{Sec:2Site} we consider the analytically solvable case of two sites and one atom. We show that numerical simulations recover the Born rule as the number of measurements $\NumMeas \rightarrow \infty$ (establishing the projective measurement limit). We further show that the measurement dynamics can be mapped to a classical diffusion equation. Section~\ref{Sec:LargeSystem} presents the main results for larger system sizes, namely the effect of measurement resolution on the collapse dynamics and steady states, the emergence of non-trivial BFSs, and how the BFSs scale with system size. We conclude in Sec.~\ref{Sec:Conclusion}.

\section{Model \label{Sec:Model}}
We consider a system of spinless, non-interacting fermions confined to a one-dimensional lattice potential. We assume the lattice is deep enough that the system Hamiltonian is well described by a tight-binding model
\begin{equation}
    H = -J\sum_j \left( \hat{c}_{j+1}^\dagger\hat{c}_{j} + \mathrm{H.c.}\right)\ .
    \label{eqn:H}
\end{equation}
Here the integer $j$ indicates  the site index (thereby measuring length in units of the lattice constant), $J$ parameterizes nearest neighbor tunneling strength, and $\hat{c}^\dagger_{j}$, $\hat{c}_{j}$ are the anticommuting fermionic creation and annihilation operators. We consider systems with periodic boundary conditions.

The ground state of Hamiltonian~\eqref{eqn:H} is the Fermi sea. The Fermi sea is described by the state $\ket{\rm FS} = \prod_{k \leq k_{\rm F}} \hat{c}^\dagger_k \ket{00\ldots 0}$, where $\hat{c}^\dagger_k$ creates a fermion of momentum $k$ and $k_{\rm F}$ indicates the Fermi momentum. The Fermi sea is bounded by the Fermi surface at $|k| = k_{\rm F}$ and is, by definition, a superposition over all possible number states. In practice, we consider systems with $M$ sites and $N$ fermions, where $N \leq M$. When $N$ is an odd number, there is a well-defined $k=0$ state and a sharp Fermi surface; the momentum distribution is $n(k) = \Theta(k_{\rm F}-|k|)$ where $\Theta$ is the Heaviside function. In one dimension the Fermi `surface' is simply a pair of zero-dimensional points at $k = \pm k_{\rm F}$. We note that the system is always an eigenstate of the total number operator $\hat N$ as we do not consider fluctuations in the total number of particles.

The system can be weakly measured using phase-contrast imaging, in which the atoms interact with a far-detuned beam of laser light of wavelength $\lambda$. A single measurement of the many-body fermionic state $\ket{\Psi}$ is described by the application of a Kraus operator $\hat{K}$, leading to the post-measurement state $|\Psi'\rangle = \hat{K}|\Psi\rangle$~\cite{Kraus1974}. The Kraus operator for a specific measurement process can be derived from the light-matter interaction Hamiltonian~\cite{hurst2019measurement, hurst2020feedback, Yamaguchi2022}.

For a phase-contrast measurement with perfect single-site spatial resolution, the unnormalized Kraus operator can be expressed as
\begin{align}
\hat K &= \exp\left[-\frac{\varphi^2}{2}\sum_j\left(\hat n_j - n_j\right)^2 \right] \nonumber \\
&= \exp\left[-\frac{\varphi^2}{2}\sum_k\left|\hat \nt_k -\nt_k\right|^2 \right], \label{Eq:Kraus}
\end{align}
in terms of coordinate- or momentum-space operators, respectively.
Here $\hat{n}_j = \hat{c}^\dagger_j \hat{c}_j$ is the usual spatial number operator, however, 
\begin{equation}
    \hat{\tilde{n}}_k =  \frac{1}{\sqrt{M}}\sum_{j =0}^{M-1}\hat{n}_j e^{-i k j},
    \label{eqn:results:Projective-Measurement-Ops}
\end{equation}
is the Fourier transform of $\hat n_j$ (not the momentum space number operator $c^\dagger_k c_k$).
In either case, $\hat K$ has a Gaussian form with width governed by $\varphi$, the dimensionless measurement strength. The quantity $n_j = \langle \hat n_j\rangle + m_j / \varphi$, a real number, indicates the coordinate-space measurement outcome for site $j$. The random variable $m_j$ quantifies projection noise with zero average ($\overline{m_j} = 0$) and no spatial correlations ($\overline{m_j m_{j'}} = \delta_{jj'}/2$)~(In this work we use $\overline{~~\cdot~~}$ to denote a statistical average and $\langle\cdot\rangle$ to denote a quantum-mechanical expectation value.). This implies Fourier domain statistics $\overline{\tilde{m}_k} = 0$, and $\overline{\tilde{m}_k\tilde{m}^*_{k'}} = \delta_{kk'}/2$, where $\tilde{m}_k$ denotes the Fourier transform of $m_j$. Thus, the Kraus operator equivalently describes measurements of the operators $\hat{\tilde{n}}_k$, with strength $\varphi$ and measurement outcomes $\tilde n_k = \langle\hat\nt_k\rangle + \tilde m_k / \varphi$.

In reality the physical measurement process cannot---even in principle---access information on a length scale smaller than $k_{\rm c}^{-1}\approx\lambda/(2\pi)$.
This physical cutoff can be expressed as a momentum-dependent measurement strength $\varphi \rightarrow \varphi \sqrt{f_k}$ where $f_k$ is a filtering function with $f_0 = 1$. In this work we model measurements with a hard cutoff at $k_{\rm c}$ given by $f_k = \Theta(k_{\rm c} - |k|)$.
Thus, the operators $\hat\nt_k$ are not measured for $k > k_{\rm c}$; in real space, the resulting backaction and measurement outcomes are correlated below the $\sim 1/k_{\rm c}$ length scale.

After implementing a cutoff, the Fourier space Kraus operator is
\begin{align}
\hat K &= \exp\left[-\frac{\varphi^2}{2}\sum_k\left|\sqrt{f_k} \left(\hat \nt_k -\<\hat \nt_k\>\right) - \frac{\tilde{m}_k}{\varphi}\right|^2 \right] \label{eq:main:Kraus_k_partial};
\end{align}
a detailed derivation connecting Eq.~\eqref{Eq:Kraus} to \eqref{eq:main:Kraus_k_partial} is provided in Appendix~\ref{App:Fourier}. This expression shows that when only the $k=0$ state is measured this problem reduces to a measurement of the total number since $\hat\nt_0 \propto \hat N$.  Since we work in system with a fixed atom number, this corresponds to a quantum nondemolition measurement, i.e. a measurement that does not impart any backaction onto the system. 

The central aim of this work is to explore the effect of the cutoff $k_{\rm c}$ on the collapse dynamics and steady-state properties of a many-body fermionic state subject to consecutive weak measurements. We investigate the collapse dynamics between two limits of the measurement resolution, where $k_{\rm c} = 0$ measures only $\hat{N}$ and $k_{\rm c}=\pi$ corresponds to no resolution cutoff.
Although a ground state Fermi sea has no momentum components above the Fermi momentum $k_{\rm F}$, spatially resolved measurements readily populate these states.

The weak measurement process described by repeated application of the Kraus operator causes a pure state to evolve along a stochastic trajectory, eventually collapsing into an eigenstate of the observable being measured~\cite{Jacobs2006}. For example, with no cutoff ($k_{\rm c} = \pi)$, repeated measurements steer the many-body fermionic system toward eigenstates of $\hat{n}_j$, denoted $\ket{{\bf n}}$ in the Fock basis, where each ${\bf n}$ corresponds to a specific spatial configuration of fermions, for example $\left(0110\ldots1\right)$. The Hamiltonian \eqref{eqn:H} is not diagonal in the Fock basis, so our initial state---the ground state---is not a Fock state. It is not clear how the collapse process proceeds even starting from the simplest ground state: the Fermi sea. This starting point will be the basis of our investigations. 

We do not consider the effects of unitary time evolution [e.g. from Eq.~\eqref{eqn:H}]  between measurements, i.e. we assume the measurement rate is much faster than the timescale $\sim \hbar/J$ for internal dynamics. We map out the dynamics of state collapse over many measurements and examine the final state of the system, i.e. the state that would be determined from a single projective measurement. 

\subsection{Numerical Implementation}

We approach this problem with a combination of analytic and numerical approaches. To avoid any dependence on an approximation scheme, the numerical approach exactly represents states in the full Hilbert space. The measurement operators $\hat{K}$ are approximately local in coordinate space so we express the many-body wavefunction as
\begin{align}
\ket{\Psi} &= \sum_{\bf n} \psi_{\bf n}\ket{\bf n} \label{eq:manybody_psi}
\end{align}
in the Fock basis of dimension $D = \rm dim(\left\{\ket{\vec{n}} \right\})$.
Our simulations record the measurement outcomes $n_j = \langle \hat{n}_j \rangle + m_j/\varphi$, the wavefunction after each measurement, and the final wavefunction after the system reaches a steady state. 

\begin{figure*}[htb!]
    \centering
    \includegraphics[width=\textwidth]{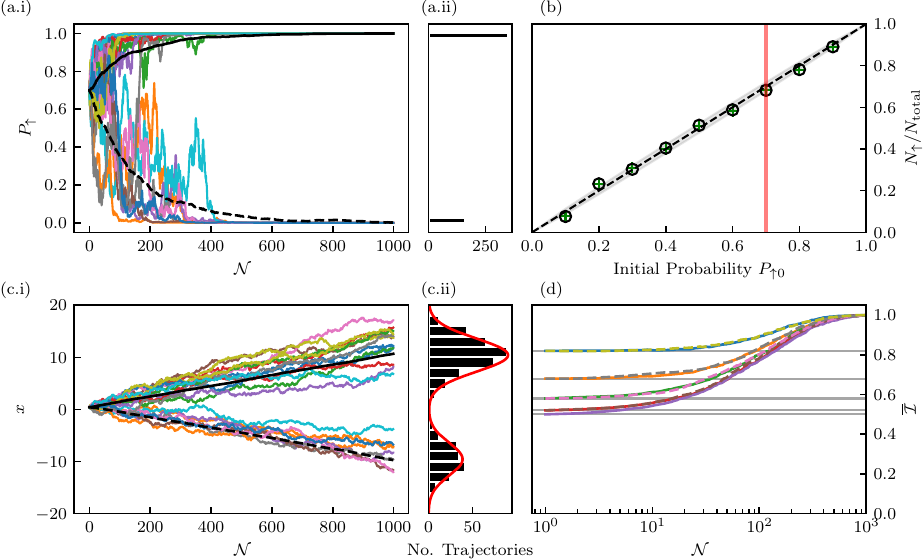}
    \caption{
    Emergence of the Born rule for $512$ trajectories after $1000$ weak measurements. 
    (a.i) Trajectories starting with $P_{\uparrow 0} = 0.7$ show  $P_\uparrow$ evolving as a function of the number of measurements; for clarity, only 24 trajectories are shown. The thick curves show the average for all trajectories, grouped by the final state assignment.
    (a.ii) Histogram of final values for all trajectories. 
    (b) Fraction of trajectories with final assignment $\ket{\uparrow}$ as a function of the initial probability $P_{\uparrow 0}$ potted along with the Born rule prediction (dashed line). 
    The grey band indicates the single-sigma statistical uncertainty associated with the $512$ trajectories. 
    Symbols compare simulations directly using Eq.~\eqref{eq:kraus_two_site} (green crosses) with diffusion model simulations from Eq.~\eqref{Eqn:update1} (black circles).
    The pink vertical line marks $P_{\uparrow 0} = 0.7$ as used in (a,c). 
    (c) Diffusion model trajectories for the same parameters as in (a).
    In (c.ii) the red curve plots the expected distribution for stochastic variable $x$ as defined in the text.
    (d) $\IPRavg$ computed for the same values of $P_{\uparrow 0}$ as in (b); gray horizontal lines mark the $\IPR$ of the initial superposition state.
    \label{fig:BornRule-ED}
    }
\end{figure*}

Equation~\eqref{eq:main:Kraus_k_partial} is numerically expensive to implement even for modest system sizes, both in terms of CPU time (computation of the matrix exponential), and storage (from the size of the Hilbert space). The Kraus operator can be approximated in real space using a series expansion 
\begin{equation}
\hat K \approx \hat{\mathbb{I}} + \varphi \sum_j \eta_j\delta\hat n_j - \frac{\varphi^2}{4} \sum_{j,j^\prime} \frac{f_{j-j^\prime}}{\sqrt{M}} \delta \hat{n}_j \delta \hat{n}_{j^\prime},  \label{eq:real_space_ensemble}
\end{equation}
where we introduced the relative density operator $\delta\hat{n}_j = \hat{n}_j - \langle \hat{n}_j\rangle$ and the identity matrix $\hat{\mathbb{I}}$. The momentum-space filtering function $f_k$ leads to a new effective noise term $\eta_j$ with statistics $\overline{\eta_j} = 0$ and
\begin{equation}
    \overline{\eta_j\eta_{j'}} = f_{j-j^\prime} = \frac{1}{\sqrt{M}}\sum_k e^{i(j-j^\prime)k}f_k. 
    \label{Eqn:newnoise}
\end{equation}
In deriving equation~\eqref{eq:real_space_ensemble} we replaced the second-order noise terms with their statistical average, i.e. $\eta_j\eta_{j'} \rightarrow \overline{\eta_j\eta_{j'}}$ this corresponds to the standard replacement of products of Wiener increments with their variance in stochastic calculus (for finite intervals this is an approximation); full details are provided in Appendix~\ref{App:Fourier}. Wavefunction evolution under repeated weak measurements is therefore completely described by successive application of the Kraus operator, i.e., after $\ell$ measurements $\ket{\Psi_\ell} = \hat{K}_\ell\ldots\hat{K}_2\hat{K}_1\ket{\mathrm{FS}}$. Note that since each $\hat{K}$ depends on the expectation values $\langle \hat{n}_j\rangle$, the Kraus operator must be calculated after each measurement.

\section{Minimal Two Site Model \label{Sec:2Site}}

We begin by benchmarking our numerical approach against the analytically solvable case of two sites with one atom. In what follows, we first identify a basis for the resulting vector space and express the number operators in that basis. This gives a direct matrix representation of the Kraus operator from which we obtain a stochastic equation of motion.

The orthonormal states $\hat c^\dagger_0\ket{00} = \ket{10} \equiv \ket{\downarrow}$ and $\hat c^\dagger_1\ket{00} = \ket{01} \equiv \ket{\uparrow}$ fully span the one-atom two-site Hilbert space.
In this basis the number operators
\begin{align}
\hat n_0 &= \frac{1}{2}\left(\hat I - \hat \sigma_z\right), & {\rm and} && \hat n_1 &= \frac{1}{2}\left(\hat I + \hat \sigma_z\right)
\end{align}
can be written in terms of the Pauli operators $\hat \sigma_{x,y,z}$ and the $2 \times 2$ identity $\hat I$.
In this representation, the Kraus operator in Eq.~\eqref{Eq:Kraus} becomes
\begin{equation}
\hat K  = \exp\left(\varphi^2\sum_j \hat n_j n_j\right)
\end{equation}
where we used the identity $\sum_{j=0}^1\hat n_j^2 = \hat I$; this simplifies to 
\begin{align}
\hat K &\propto \cosh\left(\frac{\varphi^2 \Delta n}{2}\right) \hat I  + \sinh\left(\frac{\varphi^2 \Delta n}{2}\right) \hat \sigma_z.\label{eq:kraus_two_site}
\end{align}
As expected, $\hat{K}$ depends only on the measured number difference $\Delta n = \langle\hat n_1\rangle-\langle \hat n_0\rangle + \Delta m/\varphi$, where $\Delta m \equiv m_0 - m_1$ is a new random variable with variance $1$.

To see how this changes a generic state in Hilbert space we consider $\ket{\Psi} = \cos(\theta/2)\ket{\uparrow} + \sin(\theta/2)\ket{\downarrow}$, where we omit the azimuthal angle $\phi$ since $\hat K$ leaves it unchanged.
The un-normalized post-measurement state is 
\begin{align*}
\hat K \ket{\Psi} &= e^{\varphi^2 \Delta n /2} \cos\left(\frac{\theta}{2}\right) \ket{\uparrow} + e^{-\varphi^2 \Delta n/2} \sin\left(\frac{\theta}{2}\right)\ket{\downarrow}.
\end{align*}
We use the ratio of these amplitudes to obtain the polar angle $\theta$ after one measurement
\begin{align}
\tan \left(\frac{\theta^\prime}{2}\right) &= e^{-\varphi^2 \Delta n} \tan \left(\frac{\theta}{2}\right).
\label{Eqn:thetawalk}
\end{align}
We now introduce a mapping 
\begin{align}
x = \ln[\tan (\theta/2)]\label{eq:log_transform}
\end{align}
to re-cast the measurement dynamics as a random walk for a classical stochastic variable $x$, which runs from $x\rightarrow-\infty$ for $\theta \rightarrow 0$ to $x\rightarrow\infty$ as $\theta\rightarrow \pi$; with $x=0$ corresponding to $\theta = \pi/2$ and the domain $\theta\in(0,\pi)$.
Equation~\eqref{Eqn:thetawalk} now takes the form of a diffusion model with a drift term
\begin{align}
x^\prime = x + \varphi^2 \tanh x - \varphi\Delta m,
\label{Eqn:update1}
\end{align}
where we used $\langle\hat n_1\rangle-\langle \hat n_0\rangle = \cos\theta = -\tanh x$
\footnote{This discrete update model can be connected to a continuous stochastic equation as follows. To consider a measurement of duration $dt$, we redefine our measurement strength according to $\varphi \rightarrow \varphi\sqrt{dt}$.
This allows us to introduce Weiner increment $dW = \sqrt{dt}\Delta m$ with the correct statistical properties of $\overline{dW} = 0$ and $\overline{dWdW'} = \overline{\Delta m \Delta m'} dt = dt$.}.

\subsection{Numerical confirmation}

We employ these analytic relations in a direct quantum trajectories simulations of the two site measurement process.
Throughout the manuscript our simulations employ exact numerical techniques, fully expressing the initial state in the Fock basis. The ensuing dynamics are generated entirely by the repeated application of the Kraus operator in Eq.~\eqref{eq:main:Kraus_k_partial} with measurement strength fixed at $\varphi = 0.1$. 

Figure~\ref{fig:BornRule-ED}(a.i) plots the occupation probability $P_\uparrow$ of representative individual trajectories (color) along with the average (black curves) as a function of the number of measurements $\NumMeas$, which serves as a proxy for time in these results. 
The initial state for this simulation is a superposition between the two basis states $\ket{\Psi_0} = \sqrt{P_{\uparrow0}}\ket{\uparrow} + \sqrt{1-P_{\uparrow0}}\ket{\downarrow}$ with initial probability $P_{\uparrow0}=0.7$. The histograms in Fig.~\ref{fig:BornRule-ED}(a.ii) show the final distribution of probabilities, with a fraction $N_\uparrow / N_{\rm total} = 0.683(20)$  of trajectories collapsing to $\ket{\uparrow}$.

This is consistent with our expectation that after many weak measurements the state will collapse into one of the two basis states ($\ket{\uparrow}, \ket{\downarrow}$), with probabilities given by the usual Born rule. The green crosses in Fig.~\ref{fig:BornRule-ED}(b) confirm this expectation by plotting the final probabilities after 1000 measurements as a function of the occupation probability $P_{\uparrow0}$. The black dashed line plots the expectation that $N_\uparrow / N_{\rm total} = P_{\uparrow0}$ and grey band marks the expected single-sigma statistical uncertainty for $512$ trajectories.

As a second numerical benchmark, we turn to the diffusion model in Eq.~\eqref{Eqn:update1} in which the stochastic variable is a simple linear contribution to the update rule for $x$. Figure~\ref{fig:BornRule-ED}(c.i) plots the ``position" $x$ as a function of measurement number using the same random seed as in (a), thereby yielding the same measurement history. As expected we observe a 1-to-1 correspondence between traces in the two plots, however, the mapping shows its utility in that the histograms in Fig.~\ref{fig:BornRule-ED}(c.ii) are normally distributed in contrast with those of final probability in (a.ii).

For large $|x|$, where $\tanh x \rightarrow \pm 1$, Eq.~\eqref{Eqn:update1} predicts a constant drift velocity of $\pm \varphi^2$, along with a stochastic contribution that leads to a variance $\varphi^2 \NumMeas$ after many measurements. The competition between drift and diffusive spreading determines both the rate of wavefunction collapse and the final probabilities. Using these parameters the solid red curve plots the expected bimodal distribution consisting of a sum of normal distributions with no free parameters, showing excellent agreement with the simulated histograms.

A cursory inspection of (a) might indicate that the wavefunction collapse processes is complete --in the sense that ensemble averaged dynamics have ceased-- after about $600$ measurements, as the trajectories appear to be static in the plot of $P_\uparrow$. However, (c) shows that this is not the case. We confirmed that applying the transformation in Eq.~\eqref{eq:log_transform} to the trajectories in (a) recovers the full evolution in (c); this validates the numerical stability of our Kraus operator method even with very small amplitudes and at long times.

\subsection{Metric of wavefunction collapse} 

In the two site example it is straightforward to directly track the probability of finding the system in $\ket{\downarrow}$ or $\ket{\uparrow}$ during the collapse process. However, tracking probabilities is impractical in larger systems since the number of possible projective measurement outcomes---equal to the Hilbert space dimension $D$---grows exponentially with system size.

For this reason we instead quantify state collapse using a single aggregate quantity, the inverse participation ratio (IPR) often used to study localization phenomena~\cite{Brezini1992, Evers2008}:
\begin{equation}
    \IPR = \sum_{\vec{n}} |\psi_{\vec{n}}|^{4}.
    \label{Eqn:ipr}
\end{equation}
The IPR is minimized at $\IPR = 1/D$ for states that are fully delocalized in the Fock basis, and maximized at $\IPR = 1$ for each possible Fock state $\ket{\vec{n}}$. For the two-site system the IPR can be expressed as a function of the real-space occupation probabilities, $\IPR = 2 P_\uparrow (P_\uparrow-1) + 1$. As expected, this reduces to $\IPR = 1/2$ for the fully delocalized states ($P_\uparrow = 1/2$) and $\IPR = 1$ for Fock states ($P_\uparrow = 0$ or $1$).

The average evolution of the IPR under repeated weak measurement is summarized in Fig.~\ref{fig:BornRule-ED}(d) which plots $\IPRavg$ (i.e. $\IPR$ averaged over $512$ trajectories) for the same 9 initial probabilities $P_{\uparrow0}$ as in Fig.~\ref{fig:BornRule-ED}(b), the grey lines mark the expected initial $\IPR$. Because $\IPR$ is symmetric about $P_{\uparrow0}=1/2$ only 5 curves are visible. These results confirm that the IPR is a suitable metric for wavefunction collapse driven by local measurements of the atomic density.

\section{Large-system results \label{Sec:LargeSystem}}

Having established our approach with the two-site one-atom case, we now turn to state evolution and the concomitant wavefunction collapse for a larger system of $M = 16$ sites near half filling with $N = 7$ fermions. The larger system size allows us to explore the effect of the cutoff $k_{\rm c}$, introduced in Eq.~\eqref{eq:main:Kraus_k_partial}, which limits the spatial resolution of the measurement. The repeated application of the Kraus operator $\hat K \ket{\Psi} \rightarrow \ket{\Psi^\prime}$ can be understood as a stochastic map between different points in Hilbert space; we therefore define a ``collapsed'' wavefunctions as a fixed point of the map where $\hat K \ket{\Psi_f} = \ket{\Psi_f}$, up to normalization and overall phase factors. Since the operators $\hat K \propto \exp(\hat O)$ are the exponential of some observable $\hat O$ (a Hermitian operator), the fixed points are eigenvectors of the exponentiated observable. Our results show that reducing $k_{\rm c}$ profoundly influences the structure of the fixed points $\ket{\Psi_f}$ as well as increasing the number of measurements required to achieve them. 
 
It seems reasonable that the rate of wavefunction collapse decreases with decreasing $k_{\rm c}$ since the information extracted per measurement is reduced. However, it is less obvious how $k_{\rm c}$ changes the possible final states. For example, from a physical perspective it is plausible that even with decreased spatial resolution (smaller $k_{\rm c}$), the wavefunction would eventually collapse to some Fock state that could be known with a sufficiently large number of measurements, akin to ``splitting the line'' in atomic spectroscopy~\cite{Sansonetti2011}. Our results show that this is not the case. For reduced $k_{\rm c}$, many final states are superpositions of Fock states, resulting from unresolved degeneracies in the measurement operators' spectrum.

In the following sections we show that for a given measurement protocol the entire Hilbert space can be partitioned into a non-overlapping set of BFSs. Each fixed point of the measurement map resides within a single BFS. Here, we are particularly interested in the BFS resulting from degeneracies, which can be non-trivial in the sense that the BFS corresponding to a degeneracy can contain many Fock states. Thus, even after wavefunction collapse the coherence in a non-trivial BFS can be significant, leading to $\IPR < 1$. This behavior emerges for individual trajectories and also in the ensemble average. 

\subsection{State Collapse on Average: Ensemble Behavior}

We vary the cutoff from $k_{\rm c} = \pi$ (i.e., no cutoff) to $k_{\rm c} = 0^+$ (i.e., admitting only the $k=0$ mode), and explore the effect on measurement induced dynamics. For the case of ideal position-resolved measurements with $k_{\rm c} = \pi$, the state collapses to a Fock state with atoms randomly distributed throughout the lattice. By contrast, $k_{\rm c} = 0^+$ corresponds a measurement of the total number operator $\hat N$, and the state undergoes no collapse because we consider a system with a fixed total particle number.

\begin{figure}[!tbp]
    \centering
    \includegraphics{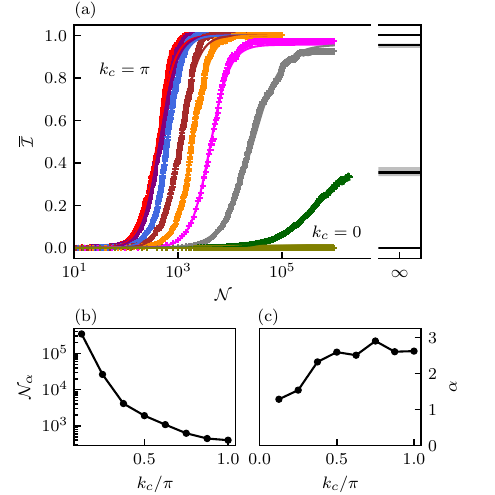}
    \caption{Dependence of average IPR on cutoff momentum for $M=16$ sites and $N=7$ fermions. (a) $\IPRavg$ as a function of number of measurements for every allowed $k_{\rm c}$.  The average was taken over 64 trajectories, except for $k_{\rm c} = 0^+$ which used 8 trajectories. The horizontal black lines indicate the computed value of $\IPRavg_\infty$ and the gray bands indicate the single-sigma statistical uncertainty for $64$ trajectories.  Solid curves are fits as described in the main text, with resulting fit
    parameters $\NumMeas_\alpha$ and $\alpha$ as a function of $k_{\rm c}$ shown in (b, c).
    \label{fig:ModelFitting}}
\end{figure}

As motivated above, Fig.~\ref{fig:ModelFitting}(a) quantifies wavefunction collapse in terms of the ensemble-averaged IPR, denoted $\IPRavg$, where each data point represents an average over 64 trajectories. For each value of $k_{\rm c}$, the IPR starts at $\IPR_0 = 5.034\times10^{-4}$, that of the initial Fermi-sea ground state~\footnote{Because the Fermi-sea ground state has spatial correlations on the scale of $1/k_{\rm F}$ it does not uniformly sample Fock space, and the associated IPR is slightly larger than zero.}, and asymptotes to a final value $\IPRavg_\infty$ as $\NumMeas\rightarrow\infty$. For $k_{\rm c} = \pi$ (red) the $\IPRavg$ rapidly rises to its asymptotic value $\IPRavg_\infty=1$, indicating that every trajectory collapses to a Fock state. The remaining traces in Fig.~\ref{fig:ModelFitting}(a) (for $k_{\rm c}<\pi$) show that while the IPR exhibits the same qualitative behavior, it increases more slowly and in some cases saturates to $\IPRavg_\infty < 1$ (implying the existence of non-trivial BFSs).

We quantify this behavior by fitting $\IPRavg$ to an empirically selected monotonically increasing function
\begin{align}
     \IPRavg(\NumMeas) &= \frac{\IPRavg_\infty}{1/f(\NumMeas) + 1} \label{eqn:results:IPR_fit} \\ 
     f(\NumMeas) &= \left(\frac{\NumMeas}{\NumMeas_\alpha}\right)^\alpha + \frac{\IPR_0}{\IPRavg_\infty-\IPR_0}, \nonumber
\end{align}
which takes on the desired limiting values.
In these fits, neither $\IPR_0$ nor $\IPRavg_\infty$ are free parameters; instead $\IPR_0$ is computed from $\ket{\rm FS}$, and $\IPRavg_{\infty}$ is determined directly from the projection of $\ket{\rm FS}$ onto the BFS, indicated by the black horizontal lines at $\NumMeas \rightarrow \infty$, which will be described further in Sec.~\ref{SubSec:QND}.
The remaining parameters $\NumMeas_\alpha$ and $\alpha$ quantify the overall shape of $\IPRavg(\NumMeas)$ during wavefunction collapse; roughly speaking $\NumMeas_\alpha$ determines when $\IPRavg$ has achieved half of its asymptotic value, and $\alpha$ captures the fractional duration of the collapse dynamics.

Figure~\ref{fig:ModelFitting}(b,c) summarizes the result of the fits for the full gamut of cutoffs. Panel (b) clearly shows the delayed onset of collapse with decreasing $k_{\rm c}$, with $\NumMeas_\alpha$ diverging as $k_{\rm c}\rightarrow0$ (at $k_{\rm c}=0^+$ the IPR is unchanged from its initial value so $\NumMeas_\alpha\rightarrow\infty$). In addition, panel (c) indicates that for large $k_{\rm c}$ the collapse dynamics is fairly abrupt (larger $\alpha$), and becomes increasingly gradual (smaller $\alpha$) for smaller $k_{\rm c}$.

These parameters give us a picture of how systems collapse on average during a weak measurement protocol. For smaller $k_{\rm c}$ the asymptotic IPR $\IPRavg_\infty$ [horizontal black lines in panel (a)] is less than one; this indicates that some fixed points $\ket{\Psi_f}$ of the Kraus operator $\hat K$ are not Fock states. To better understand this behavior, we now turn our attention to individual trajectories. 

\subsection{Emergence of backaction-free subspaces \label{SubSec:QND}}

\begin{figure}[tb!]
    \centering
    \includegraphics{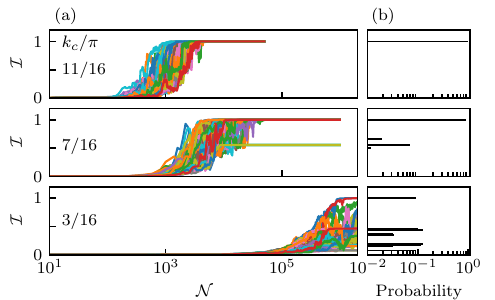}
    \caption{Behavior of individual trajectories for three values of $k_{\rm c}/\pi$.
    (a) 64 representative trajectories contributing to the averages in Fig.~\ref{fig:ModelFitting}, for  $k_{\rm c}/\pi = 11/16$, $7/16$, and $3/16$ (top to bottom).
    (b) Expected probability distribution of $\IPR_\infty$ derived from a spectral analysis of $\{\hat{\tilde{n}}_k\}_{|k| \leq k_{\rm c}}$ for the $\ket{\rm FS}$ initial state for each cutoff value in (a). 
    \label{fig:Individual Trajectories}
    }
\end{figure}

Figure~\ref{fig:Individual Trajectories} plots the individual trajectories associated with the averages shown in Fig.~\ref{fig:ModelFitting} for three cutoff values.
Individual trajectories show qualitatively similar behavior as the average, rising from $\IPR_0$ before saturating to an asymptotic value. Similar to the two-site case in Fig.~\ref{fig:BornRule-ED}, the stochastic nature of the measurement process causes some trajectories to collapse more rapidly than others. However, unlike the two-site case, individual trajectories can arrive at different asymptotic values of the IPR as shown in the left column of Fig.~\ref{fig:Individual Trajectories}. For example, for $k_{\rm c}/\pi = 7/16$ and $3/16$, some trajectories collapse to a Fock state with $\IPR = 1$, however, a significant fraction reach fixed points with $\IPR < 1$. In contrast, for $k_{\rm c}/\pi = 11/16$, all trajectories eventually collapse to a Fock state with $\IPR = 1$. In all cases, the evolution during each trajectory is stochastic, however asymptotically $\IPR$ only arrives at specific, discrete values that are determined by the cutoff.  

To further understand and quantify this behavior, we return to the Kraus operator including the cutoff $k_{\rm c}$ in Eq.~\eqref{eq:main:Kraus_k_partial}. Recall that $k_{\rm c}$ inhibits the measurement of operators with $k > k_{\rm c}$.  In our model, the operators of interest are the Fourier transformed number operators, e.g., $\hat\nt_{k}$ in Eq.~\eqref{eqn:results:Projective-Measurement-Ops}. It is straightforward to show that these operators commute and therefore form a set of compatible observables (one for each value of $k$), which is complete when every BFS has dimension $D_s=1$, i.e., there are no degeneracies. As a result, the Kraus operator can be decomposed as $\hat{K} = \prod_{|k|\leq k_{\rm c}} \hat{K}_k$  with 
\begin{equation}
    \hat{K}_k = \exp\left(-\frac{\varphi^2}{2}\left|\hat \nt_k -\<\hat \nt_k\>-\frac{\tilde{m}_k}{\varphi}\right|^2\right).
\end{equation}
Every eigenstate of $\hat\nt_k$ is an attractor of the associated Kraus map $\hat{K}_k$ (see Appendix~\ref{App:Map} for details). Furthermore, any wavefunction that is a simultaneous eigenstate of the set of measured operators $\{\hat\nt_k\}_{|k| \leq k_{\rm c}}$ is an attractor of the overall Kraus operator $\hat K$. In this sense, the BFSs are attractors and any coherences within the non-trivial BFSs are invariant under $\hat K$.

As $\NumMeas \rightarrow \infty$, each trajectory arrives at one of the attractors. In other words, the system recovers the projective measurement limit, i.e. collapse to a eigenstate of the operators being measured. In this case $\ket{\Psi_f}$ can be \emph{any} eigenstate of the measurement operators $\{\hat\nt_k\}_{|k| \leq k_{\rm c}}$. 
Once the system reaches an eigenstate, additional application of $\hat K$ no longer affects it, making subsequent measurements backaction free. In other words, with respect to the BFSs the measurement operators constitute a quantum non-demolition measurement.

Because any coherence that exists within each BFS is unchanged by $\hat K$, the measurement protocol can drive the system into a BFS where the final state is a linear combination of Fock states. Importantly, the remaining coherence is governed by whatever coherences were present in the initial state. Given an ensemble of trajectories, any trajectory that saturates to $\IPR_\infty < 1$ has been driven into a non-trivial BFS.

With this understanding, we can predict the fixed points $\ket{\Psi_f}$ of the measurement protocol by directly computing the the measurement operators' spectrum. Then, we use the projection of the Fermi-sea initial state $\ket{\mathrm{FS}}$ onto $\ket{\Psi_f}$ to obtain both the set of allowed $\IPR_\infty$ and their probabilities, shown in the histograms on the right side of Fig.~\ref{fig:Individual Trajectories}. As $k_{\rm c}$ decreases, the spectrum of $\{\hat{\tilde{n}}_k\}_{|k| \leq k_{\rm c}}$ becomes highly degenerate and the probability of the system collapsing to a non-trivial BFS becomes very large, up to $\approx 90\ \%$ for the system size studied here. 

Although these exact values of $\IPRavg_\infty$  are specific to the $\ket{\mathrm{FS}}$ initial state, they suggest our concluding question: for increasing system size at what value of $k_{\rm c}$ do BFSs with $D_s > 1$ first appear, and how do they partition the full Hilbert space?

\begin{figure}[tb!]
    \centering
    \includegraphics[width=\columnwidth]{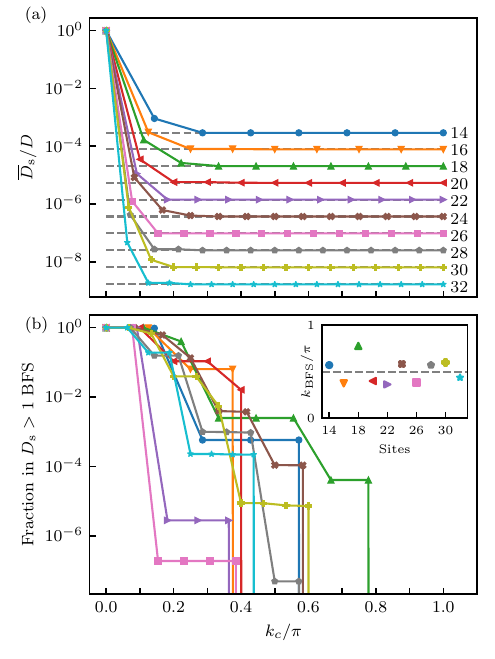}
    \caption{(a)~Average BFS dimension as a function of $k_{\rm c}/\pi$, normalized by the Hilbert space size for systems of up to $32$ sites at half filling.  The dashed horizontal lines show the asymptotic value of $1/D$, where there are no non-trivial BFS present. Numbers to the right of the traces indicate the number of sites. (b)~Fraction of states in a non-trivial BFS as a function of $k_{\rm c}$ and system size. Inset: Largest value of $k_{\rm c}$ at which at least one BFS of dimension $D_{s} > 1$ is present for different system sizes. 
    }\label{Fig:BFS-Scaling}
\end{figure}

\subsection{Scaling of Backaction Free Subspaces \label{SubSec:Scaling}}

Simulating measurement trajectories for larger system sizes is prohibitively costly due to computing time and memory limitations.
Because the measurement operators are diagonal in the Fock basis it is straightforward to \emph{exactly} compute their eigenvalues---and thereby identify the BFSs---for systems up to 32 sites, i.e. $D\sim 10^8$, before again becoming memory limited. 
This gives us generic insight into the possible endpoints of any given measurement trajectory, keeping in mind that the final distribution of fixed points is also dependent on the initial state. In this section, we obtain the eigenvalues of $\hat{\tilde{n}}_k$ for every allowed $k$ and identify the BFSs as a function of $k_{\rm c}$ and the number of sites $M$. The BFS properties also depend on the fermion number, and for brevity we focus only on half-filled systems with $N = M/2$.

We quantify the structure of the BFSs in terms of the two metrics shown in Fig.~\ref{Fig:BFS-Scaling}: (1) the average BFS dimension denoted $\overline{D}_{\rm s}$~\footnote{Let $\{\mathcal D_i\}_i$ be the set of subspaces, we define $\overline{D}_{\rm s} \equiv \overline{{\rm Dim}(\mathcal D_i)}$.}; and (2) the fraction of states in the Hilbert space that are part of a non-trivial BFS (i.e. a BFS with dimension $D_{\rm s} > 1$). Figure~\ref{Fig:BFS-Scaling}~(a) shows $\overline{D}_{\rm s} / D$ starts at $1$ for $k_{\rm c} = 0^+$ (where the entire spectrum is degenerate) and drops sharply as the cutoff momentum is increased. This drop indicates that for $k_{\rm c} > 0$ only a handful of states participate in a typical BFS. All of these examples converge to $1/D$ by $k_{\rm c}=\pi$ where every fixed point is a Fock state.

Figure~\ref{Fig:BFS-Scaling}~(b) shows that the fraction of states belonging to a BFS with dimension $D_s > 1$ can be high - even above $90\ \%$ for smaller cutoffs.  In such cases, only a small fraction of the Hilbert space is in a $D_s=1$ subspace spanned by a single Fock state.  This implies that there is a high probability that any given trajectory arrives in a non-trivial BFS at the end of the measurement protocol. For the system sizes we studied, this metric is always below $1$ for $k_{\rm c} > 0$, so some fraction of the Hilbert space resides in non-degenerate BFSs. Physically this implies that some trajectories will collapse to a Fock state even in the face of highly limited measurement resolution. 

In the specific case of $32$ sites, $16$ fermions (Hilbert space dimension $D = 601,080,390$), and cutoff $k_{\rm c}/\pi = 1/16$ (see cyan star markers in Fig.~\ref{Fig:BFS-Scaling}) $99.990\ \%$ of the Hilbert space is in a non-trivial BFS, however, a typical BFS is relatively small with average dimension of $\bar D_s = 27.9$. A typical ``resolvable'' Fock state (i.e., in a non-degenerate BFS) contains a large number of adjacent filled sites, for example $\ket{1\ldots 1 0\ldots 0}$ and $\ket{0\ldots 0 1\ldots 1}$ are extreme examples.
It is not clear how these quantities scale with increasing system size.  
For example it remains possible that, for a fixed cutoff in the thermodynamic limit, the fraction of Hilbert space in a non-trivial BFS may tend to zero: meaning that the BFS could formally exist but only in a set of measure zero.

The inset to Fig.~\ref{Fig:BFS-Scaling}~(b) plots $k_{\rm BFS}$, the maximum value of $k_{\rm c}$ for which least one BFS is non-trivial, as a function of system size. Although the Hilbert space dimension grows exponentially with system size, we find that $k_{\rm BFS}$ is nominally constant, and even appears to approach $k_{\rm BFS} \approx 0.5\pi$ for large systems.

We believe that analytic methods are the correct way to resolve both of these final questions. We spent some effort studying the measurement operators, and identified an interesting combinatoric structure. However, more work is needed to understand the consequences of this structure, particularly for larger systems that are inaccessible by the numerical methods employed here.

\section{Conclusion and Outlook \label{Sec:Conclusion}}

We established a framework for describing fermionic lattice models subject to repeated weak measurements of on-site atomic density. We showed that the spatial measurement resolution affects both the time to collapse and the final state of the system, as measured by the inverse participation ratio. 

It is somewhat surprising that these weak measurements do not always cause the system to collapse to a Fock state at the end of a trajectory. Rather, if the measurement resolution is low enough the system can also collapse into a non-trivial backaction free subspace (BFS) spanned by multiple Fock states. A single trajectory will arrive at a specific BFS stochastically, with probability given by the generalized Born rule, and therefore can still retain some coherence and entanglement.

Even without conditioning on the measurement outcomes (i.e., averaging over trajectories and assuming a generic initial state), the von Neumann entropy of the final density matrix scales as $S\sim \log(D) - \sum_s D_s\log (D_s)/D$ where $s$ indicates a sum over BFSs. We see from this expression that any BFS with $D_{\rm s} > 1$ decreases the von Neumann entropy. In our 32 site case this yields a fractional reduction of the entropy by $\approx 20\ \%$ as compared to a maximally mixed state with $S = \log(D)$.

\subsection{Experimental considerations}

On the experimental side, weak measurements with a variable resolution cutoff could be implemented using the transverse mode structure of a multi-mode optical cavity, which naturally introduces spatial filtering in the light-matter interaction. The ability to tune the resolution of atomic density measurements in this manner would allow direct exploration of how measurement-induced back-action shapes system dynamics.

A practical approach to confirming the existence of back-action free subspaces involves correlating weak measurement records with time-of-flight (TOF) distributions resulting from a 1D expansion. A realistic implementation would begin with a single spin 1D Fermi gas confined in a deep optical lattice (negligible tunneling), with size given by a longitudinal confining potential, where it would undergo a period of repeated weak measurement.

In a second phase of the experiment, the post-measurement system would undergo a 1D time-of-flight expansion with reduced lattice depth and no longitudinal confining potential, followed by a projective measurement of density. By correlating the weak measurement record with TOF distributions over many realizations, one would expect measurement outcomes associated with trivial BFSs to yield a single, well-defined TOF distribution, while those corresponding to non-trivial subspaces would exhibit a range of distinct TOF profiles.
This direct mapping between measurement trajectories and TOF outcomes provides a viable experimental route to observing BFSs in ultracold atomic systems.

\subsection{Future directions}

Taken in the perspective of measurement induced entanglement phase transitions (EPTs), the remnant entanglement is very interesting because it does not rely on any additional dynamics or entangling operations (although we note that these transitions are more often quantified by a bipartite entanglement entropy, not the von Neumann entropy). Typical EPTs occur in systems that temporally interleave partial projective measurements and entangling operations (e.g. unitary evolution or multi qubit gates)~\cite{Gullans2020, buchhold2021effective}. In contrast, the protocol studied here is more akin to a measurement-only EPT scenario~\cite{ippoliti2021entanglement}. 
Future work could investigate whether an EPT exists in the thermodynamic limit as a function of resolution cutoff or other measurement parameters which are readily accessible in experiments.

The emergence of BFSs may, in fact, be an analog of Hilbert space fragmentation~\cite{sala2020ergodicity, moudgalya2022quantum} in a monitored quantum system. 
Hilbert space fragmentation refers to Hamiltonian systems in which an exponentially large number of disconnected subspaces emerge due to dynamical constraints rather than obvious symmetries.
Similarly, in our case, the structure of BFSs is not dictated by apparent symmetries but instead arises from the spectral properties of the measurement operators.

In this context, a natural next step is to explore the impact of tunneling. In general, the tunneling Hamiltonian in Eq.~\eqref{eqn:H} will both cause transitions between BFSs and drive dynamics within individual BFSs. Because measurements drive the system into individual BFSs, combining measurements with unitary evolution could therefore lead to a quantum Zeno scenario where unitary evolution occurs primarily within a single non-trivial BFS, with the system only occasionally transitioning to another. The structure of the local hopping Hamiltonian within a BFS may be quite interesting due to this emergent measurement-induced Hilbert-space geometry. 
For example, as with conventional Hilbert space fragmentation, it is possible that some of these subspaces exhibit ergodic dynamics while others are integrable.

Finally, our results point toward the use of weak measurement for engineering new steady states of lattice fermions. Feedback could be employed to make it more likely for the system to collapse to a non-trivial BFS. Further characterization of the BFS steady states, such as their entanglement structure or nonlocal correlations, can provide insight into which BFS may be useful for applications.  One could then implement different filtering schemes to change the the measurement operators and therefore the properties of the corresponding BFSs. A major advantage of this method of quantum state engineering is that the fixed point states are unaffected by subsequent measurements, therefore enabling robust state preparation and characterization.

\begin{acknowledgments}
HMH acknowledges the support of the San Jos\'{e} State University (SJSU) Research, Scholarship, and Creative Activity assigned time program. Research activity by YHT was supported by the Division of Research and Innovation at SJSU under Award Number 23-SRA-08-040. 
HMH was supported by the National Science Foundation under Award Number PHY-2309331. 
IBS was partially supported by the National Institute of Standards and Technology (NIST), the National Science Foundation through the Quantum Leap Challenge Institute for Robust Quantum Simulation (OMA-2120757), and the Air Force Office of Scientific Research Multidisciplinary University Research Initiative `RAPSYDY in Q' (FA9550-22-1-0339). 
The content is solely the responsibility of the authors and does not necessarily represent the official views of SJSU or NIST.
\end{acknowledgments}

\appendix 

\section{Kraus operator as a stochastic map} \label{App:Map}

We consider a general Gaussian Kraus operator
\begin{align}
\hat K &= \exp\left[-\frac{\varphi^2}{2}\sum_j\left(\hat O_j - O_j\right)^2\right]
\end{align}
with measurement outcomes $O_j = \langle\hat O_j\rangle + m_j / \varphi$, measurement strength $\varphi$, measurement noise $\overline{m_j} = 0$ and $\overline{m_j m_{j^\prime}} = \delta_{j,j^\prime}/2$. We assume that the measurement operators $\hat O_j$ commute [ Eq.~\eqref{eq:main:Kraus_k_partial} is therefore a special case].
The overall operator $\hat K$ can be written as product of individual Kraus operators $\hat K_j$, one for each $\hat O_j$, which can be studied separately.
We now consider an initial state
\begin{align}
\ket{\Psi} &\approx \ket{o^\ell} + \sum_{\ell^\prime \neq \ell} \epsilon^{\ell^\prime} \ket{o^{\ell^\prime}}
\end{align}
close to an eigenstate of these operators, expressed in the eigenbasis $\left\{o^\ell\right\}$ of $\hat O$ (the subscript $j$ is suppressed for clarity).
At first order in these $\epsilon$'s we have the expectation value $\langle\hat O\rangle = o^\ell$, leading to the action of the Kraus operator
\begin{align}
\hat K \ket{\Psi} &= e^{-m^2 / 2} \ket{o^\ell} + \sum_{\ell^\prime \neq \ell} \epsilon^{\ell^\prime} e^{-\varphi^2 (o^{\ell^\prime} - o^\ell - m/\varphi)^2 / 2} \ket{o^{\ell^\prime}}.
\end{align}
We then factor out the first exponential as it serves only to changes the overall normalization; this leaves behind quantities such as
\begin{align}
\exp\left\{-\frac{\varphi^2}{2}\left[\left(o^{\ell^\prime} - o^\ell\right)^2 - 2(o^{\ell^\prime} - o^\ell) \frac{m}{\varphi}\right]\right\}.
\end{align}
To identify the average behavior of this stochastic map, we take the ensemble average by expanding the contribution of the exponential with $m$ to second order in $\varphi$.  
Renormalizing, we arrive at 
\begin{align}
\overline{\hat K \ket{\Psi}} &\rightarrow \ket{o^\ell} + \sum_{\ell^\prime \neq \ell} \epsilon^{\ell^\prime} e^{-\varphi^2 (o^{\ell^\prime} - o^\ell)^2 / 4} \ket{o^{\ell^\prime}};
\end{align}
compared to the initial state, the perturbing terms have each decreased in amplitude by an exponential factor that is equal to $1$ in the case of degeneracies.
Thus each group of degenerate eigenstates (i.e., each BFS) is an attractor of this stochastic map.

\section{Fourier measurement operators \label{App:Fourier}}

Here we consider an interpretation of the Kraus operator in Fourier space, 
\begin{align}
\hat K &= \exp\left[-\frac{\varphi^2}{2}\sum_j\left(\hat n_j -\<\hat n_j\> - \frac{m_j}{\varphi}\right)^2 \right] \nonumber \\
&= \exp\left[-\frac{\varphi^2}{2}\sum_k\left|\hat \nt_k -\<\hat \nt_k\> - \frac{m_k}{\varphi}\right|^2 \right] \label{eq:Kraus_k_full}
\end{align}
as describing measurements the Fourier transformed density operator
\begin{align}
\hat\nt_k &= \frac{1}{\sqrt{M}} \sum_{j} \hat c^\dagger_j \hat c_j e^{-i k j} = \frac{1}{\sqrt{M}} \sum_{k^\prime} \hat c^\dagger_{k+k^\prime} \hat c_{k^\prime};
\end{align}
notice that  $\hat\nt_k$ is not the same as the fermion momentum density operator $\hat n_k = \hat c^\dagger_k \hat c_k$, and is a non-Hermitian operator. 
Since $\hat\nt_k$ is diagonalizable -- just with complex eigenvalues -- we decompose it into a pair of Hermitian components $(\hat\nt_k + \hat\nt_k^\dagger)/2$ and $(\hat\nt_k - \hat\nt_k^\dagger)/(2 i)$, respectively associated with the real and imaginary parts of the eigenvalues of $\hat\nt_k$.

\subsection{Implementation of cutoff}

This description provides a natural way to implement a cutoff in terms of a $k$-dependant measurement strength $\varphi_k = \varphi \sqrt{f_k}$, where we take $f_k$ to be a real valued, non-negative symmetric function with $f(0) = 1$. For our simple cutoff, $f_k = \Theta(k_{\rm c} - |k|)$, so $f_k$ is either $0$ or $1$. The resulting Kraus operator
\begin{align}
\hat K &= \exp\left[-\frac{\varphi^2}{2}\sum_k\left|\sqrt{f_k} \left(\hat \nt_k -\<\hat \nt_k\>\right) - \frac{m_k}{\varphi}\right|^2 \right] \label{eq:Kraus_k_partial}
\end{align}
provides an equivalent physical interpretation where new operators $\sqrt{f_k}~\hat \nt_k$ are measured with strength $\varphi$. Our prescription is to expand the argument of the exponential and factor out $\sum_k |m_k|^2/2$, which, while zero-order in $\varphi$, only contributes an overall change in normalization to the wavefunction and therefore does not contribute to measurement backaction.
The remainder of the Kraus operator is 
\begin{align}
\hat K &= \exp\left[\sum_k\left(  \varphi \sqrt{f_k} \delta\hat \nt_k  m^*_k - \frac{\varphi^2 f_k}{2} \left|\delta\hat \nt_k \right|^2 \right) \right],  \label{eq:KrausKOperator}
\end{align}
which leads to meaningful changes in the wavefunction. Recall that in our notation $\delta\nt_k = \nt_k - \langle \nt_k \rangle$. We used $\delta\hat \nt_k  = \delta\hat \nt_{-k}^\dagger$, $m_k = m^*_{-k}$, and re-indexed the sum to obtain the second term. In what follows we consider the behavior of this operator in both real and momentum space.

\subsubsection{Momentum space}

We follow the prescription for the series expansion in the main body and obtain the approximate Kraus operator
\begin{align}
\hat K &\approx 1 + \varphi \sum_k \sqrt{f_k}_k \delta \hat \nt_k  m_k^*  \label{eq:FullSecondOrder}\\
& + \frac{\varphi^2}{2} \left[ \left(\sum_{k} \sqrt{f_k} \delta \hat \nt_k m^*_k\right)^2 - \sum_k f_k |\delta \hat \nt_k|^2 \right]. \nonumber 
\end{align}
The second-to-last term can be simplified 
\begin{align}
\left(\cdots\right)^2 &= \sum_{k,k^\prime} \sqrt{f_k} \sqrt{f_{k^\prime}} \delta \hat \nt_k \delta \hat \nt_{k^\prime}  m^*_k m^*_{k^\prime} \nonumber \\
&\approx \frac{1}{2} \sum_{k}  \sqrt{f_k} \sqrt{f_{-k}} \delta \hat \nt_k \delta \hat \nt_{-k} = \frac{1}{2} \sum_{k} f_k |\delta \hat \nt_k|^2 \label{eq:Krauss_second_order}
\end{align}
having replaced the terms $m^*_k m^*_{k^\prime}$ with their ensemble average, using $\overline{m_k m^*_{k^\prime}} = \delta_{k,k^\prime}/2$ and
$\overline{m_k m_{k^\prime}} = \delta_{k,-k^\prime}/2$, from $m_j$ being real valued.

All together we find the ensemble-averaged second order Kraus operator
\begin{align}
\hat K &\approx 1 + \varphi \sum_k \sqrt{f_k} \delta \hat \nt_k  m_k^* -\frac{\varphi^2}{4} \sum_k f_k |\delta \hat \nt_k|^2. \label{eq:Kraus_cutoff_series}
\end{align}
Because this expression is written in terms of commuting density operators it is valid for both fermionic and bosonic systems. The key assumptions that go into this expression are: (1) that $\varphi$ is a small parameter, i.e. the series expansion is valid; and (2) that we can replace two-noise terms $\propto m_k m^*_{k'}$ with their ensemble average, i.e. $m_k m^*_{k'} \rightarrow \overline{m_k m^*_{k'}}$.

\subsubsection{Real space}

Here we compute the Fourier transform of the two terms in Eq.~\eqref{eq:KrausKOperator}.
The first order term can be simplified by combining $\sqrt{f_k}$ and $m_k$ into an effective noise $\eta_k \equiv \sqrt{f_k} m_k$ from which we find
\begin{align}
\sum_k \delta\hat \nt_k  \eta^*_k &= \sum_j \delta\hat n_j \eta_j.
\end{align}
In real space the second order term is 
\begin{align}
\sum_k f_k \left|\delta\hat \nt_k \right|^2 &= \sum_{j,j^\prime}  \delta \hat{n}_j \delta \hat{n}^\dagger_{j^\prime} \left[\frac{1}{M}\sum_k e^{i (j-j^\prime)k} f_k\right] \nonumber\\
&= \sum_{j,j^\prime}  \delta \hat{n}_j \delta \hat{n}_{j^\prime} \left(\frac{f_{j-j^\prime}}{\sqrt{M}} \right), 
\end{align}
where $f_{j-j^\prime}$ is defined in Eq.~\eqref{Eqn:newnoise} in the main text. With these replacements, the complete expression from Eq.~\eqref{eq:FullSecondOrder} becomes
\begin{align}
\hat K &= 1 + \varphi \sum_j \delta\hat n_j \eta_j \label{eq:app:real_space} \\
&\ \ \ \ + \frac{\varphi^2}{2} \sum_{j,j^\prime} \left[  \eta_j \eta_{j^\prime} - \left(\frac{f_{j-j^\prime}}{\sqrt{M}} \right) \right] \delta \hat{n}_j \delta \hat{n}_{j^\prime}, \nonumber
\end{align}
and the expression with the ensemble average reduces to
\begin{align}
\hat K &= 1 + \varphi \sum_j \delta\hat n_j \eta_j - \frac{\varphi^2}{4} \sum_{j,j^\prime} \frac{f_{j-j^\prime}}{\sqrt{M}} \delta \hat{n}_j \delta \hat{n}_{j^\prime}. \label{eq:app:real_space_ensemble}
\end{align}

References~\cite{hurst2020feedback} and~\cite{Yamaguchi2022} implemented the cutoff in a mean-field theory by simply scaling the noise-variance of $m_k$ by $f_k$.
As can be seen from Eq.~\eqref{eq:Kraus_cutoff_series} following this procedure gives the same first order contribution, but the algebra leading to the second order term in Eq.~\eqref{eq:Krauss_second_order} differs. In a single-component mean-field setting the second order term serves only to enforce the normalization condition, and as such scaling the noise to implement a $k$-dependent measurement strength suffices. 

The relations between the different mean field descriptions, as well as the impact of taking the ensemble average or not---e.g., contrasting Eqs.~\eqref{eq:app:real_space} and \eqref{eq:app:real_space_ensemble}---are interesting questions for future study as they address the accuracy of second-order stochastic wavefunction descriptions of open quantum systems.

\bibliography{main}

\begin{thebibliography}{42}%
\makeatletter
\providecommand \@ifxundefined [1]{%
 \@ifx{#1\undefined}
}%
\providecommand \@ifnum [1]{%
 \ifnum #1\expandafter \@firstoftwo
 \else \expandafter \@secondoftwo
 \fi
}%
\providecommand \@ifx [1]{%
 \ifx #1\expandafter \@firstoftwo
 \else \expandafter \@secondoftwo
 \fi
}%
\providecommand \natexlab [1]{#1}%
\providecommand \enquote  [1]{``#1''}%
\providecommand \bibnamefont  [1]{#1}%
\providecommand \bibfnamefont [1]{#1}%
\providecommand \citenamefont [1]{#1}%
\providecommand \href@noop [0]{\@secondoftwo}%
\providecommand \href [0]{\begingroup \@sanitize@url \@href}%
\providecommand \@href[1]{\@@startlink{#1}\@@href}%
\providecommand \@@href[1]{\endgroup#1\@@endlink}%
\providecommand \@sanitize@url [0]{\catcode `\\12\catcode `\$12\catcode
  `\&12\catcode `\#12\catcode `\^12\catcode `\_12\catcode `\%12\relax}%
\providecommand \@@startlink[1]{}%
\providecommand \@@endlink[0]{}%
\providecommand \url  [0]{\begingroup\@sanitize@url \@url }%
\providecommand \@url [1]{\endgroup\@href {#1}{\urlprefix }}%
\providecommand \urlprefix  [0]{URL }%
\providecommand \Eprint [0]{\href }%
\providecommand \doibase [0]{https://doi.org/}%
\providecommand \selectlanguage [0]{\@gobble}%
\providecommand \bibinfo  [0]{\@secondoftwo}%
\providecommand \bibfield  [0]{\@secondoftwo}%
\providecommand \translation [1]{[#1]}%
\providecommand \BibitemOpen [0]{}%
\providecommand \bibitemStop [0]{}%
\providecommand \bibitemNoStop [0]{.\EOS\space}%
\providecommand \EOS [0]{\spacefactor3000\relax}%
\providecommand \BibitemShut  [1]{\csname bibitem#1\endcsname}%
\let\auto@bib@innerbib\@empty
\bibitem [{\citenamefont {Diehl}\ \emph {et~al.}(2008)\citenamefont {Diehl},
  \citenamefont {Micheli}, \citenamefont {Kantian}, \citenamefont {Kraus},
  \citenamefont {B{\"u}chler},\ and\ \citenamefont {Zoller}}]{Diehl2008}%
  \BibitemOpen
  \bibfield  {author} {\bibinfo {author} {\bibfnamefont {S.}~\bibnamefont
  {Diehl}}, \bibinfo {author} {\bibfnamefont {A.}~\bibnamefont {Micheli}},
  \bibinfo {author} {\bibfnamefont {A.}~\bibnamefont {Kantian}}, \bibinfo
  {author} {\bibfnamefont {B.}~\bibnamefont {Kraus}}, \bibinfo {author}
  {\bibfnamefont {H.}~\bibnamefont {B{\"u}chler}},\ and\ \bibinfo {author}
  {\bibfnamefont {P.}~\bibnamefont {Zoller}},\ }\bibfield  {title} {\bibinfo
  {title} {Quantum states and phases in driven open quantum systems with cold
  atoms},\ }\href@noop {} {\bibfield  {journal} {\bibinfo  {journal} {Nature
  Physics}\ }\textbf {\bibinfo {volume} {4}},\ \bibinfo {pages} {878} (\bibinfo
  {year} {2008})}\BibitemShut {NoStop}%
\bibitem [{\citenamefont {Skinner}\ \emph {et~al.}(2019)\citenamefont
  {Skinner}, \citenamefont {Ruhman},\ and\ \citenamefont
  {Nahum}}]{skinner2019measurement}%
  \BibitemOpen
  \bibfield  {author} {\bibinfo {author} {\bibfnamefont {B.}~\bibnamefont
  {Skinner}}, \bibinfo {author} {\bibfnamefont {J.}~\bibnamefont {Ruhman}},\
  and\ \bibinfo {author} {\bibfnamefont {A.}~\bibnamefont {Nahum}},\ }\bibfield
   {title} {\bibinfo {title} {Measurement-induced phase transitions in the
  dynamics of entanglement},\ }\href@noop {} {\bibfield  {journal} {\bibinfo
  {journal} {Physical Review X}\ }\textbf {\bibinfo {volume} {9}},\ \bibinfo
  {pages} {031009} (\bibinfo {year} {2019})}\BibitemShut {NoStop}%
\bibitem [{\citenamefont {Gullans}\ and\ \citenamefont
  {Huse}(2020)}]{Gullans2020}%
  \BibitemOpen
  \bibfield  {author} {\bibinfo {author} {\bibfnamefont {M.~J.}\ \bibnamefont
  {Gullans}}\ and\ \bibinfo {author} {\bibfnamefont {D.~A.}\ \bibnamefont
  {Huse}},\ }\bibfield  {title} {\bibinfo {title} {Dynamical purification phase
  transition induced by quantum measurements},\ }\href@noop {} {\bibfield
  {journal} {\bibinfo  {journal} {Phys. Rev. X}\ }\textbf {\bibinfo {volume}
  {10}},\ \bibinfo {pages} {041020} (\bibinfo {year} {2020})}\BibitemShut
  {NoStop}%
\bibitem [{\citenamefont {Lavasani}\ \emph {et~al.}(2021)\citenamefont
  {Lavasani}, \citenamefont {Alavirad},\ and\ \citenamefont
  {Barkeshli}}]{lavasani2021measurement}%
  \BibitemOpen
  \bibfield  {author} {\bibinfo {author} {\bibfnamefont {A.}~\bibnamefont
  {Lavasani}}, \bibinfo {author} {\bibfnamefont {Y.}~\bibnamefont {Alavirad}},\
  and\ \bibinfo {author} {\bibfnamefont {M.}~\bibnamefont {Barkeshli}},\
  }\bibfield  {title} {\bibinfo {title} {Measurement-induced topological
  entanglement transitions in symmetric random quantum circuits},\ }\href@noop
  {} {\bibfield  {journal} {\bibinfo  {journal} {Nature Physics}\ }\textbf
  {\bibinfo {volume} {17}},\ \bibinfo {pages} {342} (\bibinfo {year}
  {2021})}\BibitemShut {NoStop}%
\bibitem [{\citenamefont {Ippoliti}\ \emph {et~al.}(2021)\citenamefont
  {Ippoliti}, \citenamefont {Gullans}, \citenamefont {Gopalakrishnan},
  \citenamefont {Huse},\ and\ \citenamefont
  {Khemani}}]{ippoliti2021entanglement}%
  \BibitemOpen
  \bibfield  {author} {\bibinfo {author} {\bibfnamefont {M.}~\bibnamefont
  {Ippoliti}}, \bibinfo {author} {\bibfnamefont {M.~J.}\ \bibnamefont
  {Gullans}}, \bibinfo {author} {\bibfnamefont {S.}~\bibnamefont
  {Gopalakrishnan}}, \bibinfo {author} {\bibfnamefont {D.~A.}\ \bibnamefont
  {Huse}},\ and\ \bibinfo {author} {\bibfnamefont {V.}~\bibnamefont
  {Khemani}},\ }\bibfield  {title} {\bibinfo {title} {Entanglement phase
  transitions in measurement-only dynamics},\ }\href@noop {} {\bibfield
  {journal} {\bibinfo  {journal} {Physical Review X}\ }\textbf {\bibinfo
  {volume} {11}},\ \bibinfo {pages} {011030} (\bibinfo {year}
  {2021})}\BibitemShut {NoStop}%
\bibitem [{\citenamefont {Briegel}\ \emph {et~al.}(2009)\citenamefont
  {Briegel}, \citenamefont {Browne}, \citenamefont {D{\"u}r}, \citenamefont
  {Raussendorf},\ and\ \citenamefont {Van~den Nest}}]{Briegel2009}%
  \BibitemOpen
  \bibfield  {author} {\bibinfo {author} {\bibfnamefont {H.~J.}\ \bibnamefont
  {Briegel}}, \bibinfo {author} {\bibfnamefont {D.~E.}\ \bibnamefont {Browne}},
  \bibinfo {author} {\bibfnamefont {W.}~\bibnamefont {D{\"u}r}}, \bibinfo
  {author} {\bibfnamefont {R.}~\bibnamefont {Raussendorf}},\ and\ \bibinfo
  {author} {\bibfnamefont {M.}~\bibnamefont {Van~den Nest}},\ }\bibfield
  {title} {\bibinfo {title} {Measurement-based quantum computation},\
  }\href@noop {} {\bibfield  {journal} {\bibinfo  {journal} {Nature Physics}\
  }\textbf {\bibinfo {volume} {5}},\ \bibinfo {pages} {19} (\bibinfo {year}
  {2009})}\BibitemShut {NoStop}%
\bibitem [{\citenamefont {Andrews}\ \emph {et~al.}(1997)\citenamefont
  {Andrews}, \citenamefont {Kurn}, \citenamefont {Miesner}, \citenamefont
  {Durfee}, \citenamefont {Townsend}, \citenamefont {Inouye},\ and\
  \citenamefont {Ketterle}}]{andrews1997propagation}%
  \BibitemOpen
  \bibfield  {author} {\bibinfo {author} {\bibfnamefont {M.~R.}\ \bibnamefont
  {Andrews}}, \bibinfo {author} {\bibfnamefont {D.~M.}\ \bibnamefont {Kurn}},
  \bibinfo {author} {\bibfnamefont {H.-J.}\ \bibnamefont {Miesner}}, \bibinfo
  {author} {\bibfnamefont {D.~S.}\ \bibnamefont {Durfee}}, \bibinfo {author}
  {\bibfnamefont {C.~G.}\ \bibnamefont {Townsend}}, \bibinfo {author}
  {\bibfnamefont {S.}~\bibnamefont {Inouye}},\ and\ \bibinfo {author}
  {\bibfnamefont {W.}~\bibnamefont {Ketterle}},\ }\bibfield  {title} {\bibinfo
  {title} {Propagation of sound in a bose-einstein condensate},\ }\href@noop {}
  {\bibfield  {journal} {\bibinfo  {journal} {Physical review letters}\
  }\textbf {\bibinfo {volume} {79}},\ \bibinfo {pages} {553} (\bibinfo {year}
  {1997})}\BibitemShut {NoStop}%
\bibitem [{\citenamefont {Ketterle}\ \emph {et~al.}(1999)\citenamefont
  {Ketterle}, \citenamefont {Durfee},\ and\ \citenamefont
  {Stamper-Kurn}}]{Ketterle1999}%
  \BibitemOpen
  \bibfield  {author} {\bibinfo {author} {\bibfnamefont {W.}~\bibnamefont
  {Ketterle}}, \bibinfo {author} {\bibfnamefont {D.~S.}\ \bibnamefont
  {Durfee}},\ and\ \bibinfo {author} {\bibfnamefont {D.~M.}\ \bibnamefont
  {Stamper-Kurn}},\ }\bibinfo {title} {{Making, probing and understanding
  Bose-Einstein condensates}}\ (\bibinfo  {publisher} {IOS Press},\ \bibinfo
  {year} {1999})\ pp.\ \bibinfo {pages} {67 -- 176}\BibitemShut {NoStop}%
\bibitem [{\citenamefont {Shin}\ \emph {et~al.}(2006)\citenamefont {Shin},
  \citenamefont {Zwierlein}, \citenamefont {Schunck}, \citenamefont
  {Schirotzek},\ and\ \citenamefont {Ketterle}}]{shin2006observation}%
  \BibitemOpen
  \bibfield  {author} {\bibinfo {author} {\bibfnamefont {Y.}~\bibnamefont
  {Shin}}, \bibinfo {author} {\bibfnamefont {M.~W.}\ \bibnamefont {Zwierlein}},
  \bibinfo {author} {\bibfnamefont {C.~H.}\ \bibnamefont {Schunck}}, \bibinfo
  {author} {\bibfnamefont {A.}~\bibnamefont {Schirotzek}},\ and\ \bibinfo
  {author} {\bibfnamefont {W.}~\bibnamefont {Ketterle}},\ }\bibfield  {title}
  {\bibinfo {title} {Observation of phase separation in a strongly interacting
  imbalanced fermi gas},\ }\href
  {https://doi.org/10.1103/PhysRevLett.97.030401} {\bibfield  {journal}
  {\bibinfo  {journal} {Phys. Rev. Lett.}\ }\textbf {\bibinfo {volume} {97}},\
  \bibinfo {pages} {030401} (\bibinfo {year} {2006})}\BibitemShut {NoStop}%
\bibitem [{\citenamefont {Shin}\ \emph {et~al.}(2007)\citenamefont {Shin},
  \citenamefont {Schunck}, \citenamefont {Schirotzek},\ and\ \citenamefont
  {Ketterle}}]{shin2007tomographic}%
  \BibitemOpen
  \bibfield  {author} {\bibinfo {author} {\bibfnamefont {Y.-I.}\ \bibnamefont
  {Shin}}, \bibinfo {author} {\bibfnamefont {C.}~\bibnamefont {Schunck}},
  \bibinfo {author} {\bibfnamefont {A.}~\bibnamefont {Schirotzek}},\ and\
  \bibinfo {author} {\bibfnamefont {W.}~\bibnamefont {Ketterle}},\ }\bibfield
  {title} {\bibinfo {title} {Tomographic rf spectroscopy of a trapped fermi gas
  at unitarity},\ }\href@noop {} {\bibfield  {journal} {\bibinfo  {journal}
  {Physical review letters}\ }\textbf {\bibinfo {volume} {99}},\ \bibinfo
  {pages} {090403} (\bibinfo {year} {2007})}\BibitemShut {NoStop}%
\bibitem [{\citenamefont {Altunta{\c s}}\ and\ \citenamefont
  {Spielman}(2023)}]{Altuntas2023}%
  \BibitemOpen
  \bibfield  {author} {\bibinfo {author} {\bibfnamefont {E.}~\bibnamefont
  {Altunta{\c s}}}\ and\ \bibinfo {author} {\bibfnamefont {I.~B.}\ \bibnamefont
  {Spielman}},\ }\bibfield  {title} {\bibinfo {title} {Quantum back-action
  limits in dispersively measured bose-einstein condensates},\ }\href@noop {}
  {\bibfield  {journal} {\bibinfo  {journal} {Communications Physics}\ }\textbf
  {\bibinfo {volume} {6}},\ \bibinfo {pages} {66} (\bibinfo {year}
  {2023})}\BibitemShut {NoStop}%
\bibitem [{\citenamefont {Iadecola}\ \emph {et~al.}(2023)\citenamefont
  {Iadecola}, \citenamefont {Ganeshan}, \citenamefont {Pixley},\ and\
  \citenamefont {Wilson}}]{iadecola2023measurement}%
  \BibitemOpen
  \bibfield  {author} {\bibinfo {author} {\bibfnamefont {T.}~\bibnamefont
  {Iadecola}}, \bibinfo {author} {\bibfnamefont {S.}~\bibnamefont {Ganeshan}},
  \bibinfo {author} {\bibfnamefont {J.}~\bibnamefont {Pixley}},\ and\ \bibinfo
  {author} {\bibfnamefont {J.~H.}\ \bibnamefont {Wilson}},\ }\bibfield  {title}
  {\bibinfo {title} {Measurement and feedback driven entanglement transition in
  the probabilistic control of chaos},\ }\href@noop {} {\bibfield  {journal}
  {\bibinfo  {journal} {Physical review letters}\ }\textbf {\bibinfo {volume}
  {131}},\ \bibinfo {pages} {060403} (\bibinfo {year} {2023})}\BibitemShut
  {NoStop}%
\bibitem [{\citenamefont {Agrawal}\ \emph {et~al.}(2022)\citenamefont
  {Agrawal}, \citenamefont {Zabalo}, \citenamefont {Chen}, \citenamefont
  {Wilson}, \citenamefont {Potter}, \citenamefont {Pixley}, \citenamefont
  {Gopalakrishnan},\ and\ \citenamefont {Vasseur}}]{Agrawal2022}%
  \BibitemOpen
  \bibfield  {author} {\bibinfo {author} {\bibfnamefont {U.}~\bibnamefont
  {Agrawal}}, \bibinfo {author} {\bibfnamefont {A.}~\bibnamefont {Zabalo}},
  \bibinfo {author} {\bibfnamefont {K.}~\bibnamefont {Chen}}, \bibinfo {author}
  {\bibfnamefont {J.~H.}\ \bibnamefont {Wilson}}, \bibinfo {author}
  {\bibfnamefont {A.~C.}\ \bibnamefont {Potter}}, \bibinfo {author}
  {\bibfnamefont {J.}~\bibnamefont {Pixley}}, \bibinfo {author} {\bibfnamefont
  {S.}~\bibnamefont {Gopalakrishnan}},\ and\ \bibinfo {author} {\bibfnamefont
  {R.}~\bibnamefont {Vasseur}},\ }\bibfield  {title} {\bibinfo {title}
  {Entanglement and charge-sharpening transitions in u (1) symmetric monitored
  quantum circuits},\ }\href@noop {} {\bibfield  {journal} {\bibinfo  {journal}
  {Physical Review X}\ }\textbf {\bibinfo {volume} {12}},\ \bibinfo {pages}
  {041002} (\bibinfo {year} {2022})}\BibitemShut {NoStop}%
\bibitem [{\citenamefont {Ma}\ \emph {et~al.}(2018)\citenamefont {Ma},
  \citenamefont {Wang}, \citenamefont {Leong},\ and\ \citenamefont
  {Liu}}]{Ma2018}%
  \BibitemOpen
  \bibfield  {author} {\bibinfo {author} {\bibfnamefont {W.-L.}\ \bibnamefont
  {Ma}}, \bibinfo {author} {\bibfnamefont {P.}~\bibnamefont {Wang}}, \bibinfo
  {author} {\bibfnamefont {W.-H.}\ \bibnamefont {Leong}},\ and\ \bibinfo
  {author} {\bibfnamefont {R.-B.}\ \bibnamefont {Liu}},\ }\bibfield  {title}
  {\bibinfo {title} {Phase transitions in sequential weak measurements},\
  }\href {https://doi.org/10.1103/PhysRevA.98.012117} {\bibfield  {journal}
  {\bibinfo  {journal} {Phys. Rev. A}\ }\textbf {\bibinfo {volume} {98}},\
  \bibinfo {pages} {012117} (\bibinfo {year} {2018})}\BibitemShut {NoStop}%
\bibitem [{\citenamefont {Garc{\'\i}a-Pintos}\ \emph
  {et~al.}(2019)\citenamefont {Garc{\'\i}a-Pintos}, \citenamefont {Tielas},\
  and\ \citenamefont {Del~Campo}}]{Garcia2019}%
  \BibitemOpen
  \bibfield  {author} {\bibinfo {author} {\bibfnamefont {L.~P.}\ \bibnamefont
  {Garc{\'\i}a-Pintos}}, \bibinfo {author} {\bibfnamefont {D.}~\bibnamefont
  {Tielas}},\ and\ \bibinfo {author} {\bibfnamefont {A.}~\bibnamefont
  {Del~Campo}},\ }\bibfield  {title} {\bibinfo {title} {Spontaneous symmetry
  breaking induced by quantum monitoring},\ }\href@noop {} {\bibfield
  {journal} {\bibinfo  {journal} {Physical Review Letters}\ }\textbf {\bibinfo
  {volume} {123}},\ \bibinfo {pages} {090403} (\bibinfo {year}
  {2019})}\BibitemShut {NoStop}%
\bibitem [{\citenamefont {Mu{\~n}oz-Arias}\ \emph {et~al.}(2020)\citenamefont
  {Mu{\~n}oz-Arias}, \citenamefont {Deutsch}, \citenamefont {Jessen},\ and\
  \citenamefont {Poggi}}]{munoz2020simulation}%
  \BibitemOpen
  \bibfield  {author} {\bibinfo {author} {\bibfnamefont {M.~H.}\ \bibnamefont
  {Mu{\~n}oz-Arias}}, \bibinfo {author} {\bibfnamefont {I.~H.}\ \bibnamefont
  {Deutsch}}, \bibinfo {author} {\bibfnamefont {P.~S.}\ \bibnamefont
  {Jessen}},\ and\ \bibinfo {author} {\bibfnamefont {P.~M.}\ \bibnamefont
  {Poggi}},\ }\bibfield  {title} {\bibinfo {title} {Simulation of the complex
  dynamics of mean-field p-spin models using measurement-based quantum feedback
  control},\ }\href@noop {} {\bibfield  {journal} {\bibinfo  {journal}
  {Physical Review A}\ }\textbf {\bibinfo {volume} {102}},\ \bibinfo {pages}
  {022610} (\bibinfo {year} {2020})}\BibitemShut {NoStop}%
\bibitem [{\citenamefont {Lang}\ and\ \citenamefont
  {B{\"u}chler}(2020)}]{lang2020entanglement}%
  \BibitemOpen
  \bibfield  {author} {\bibinfo {author} {\bibfnamefont {N.}~\bibnamefont
  {Lang}}\ and\ \bibinfo {author} {\bibfnamefont {H.~P.}\ \bibnamefont
  {B{\"u}chler}},\ }\bibfield  {title} {\bibinfo {title} {Entanglement
  transition in the projective transverse field ising model},\ }\href@noop {}
  {\bibfield  {journal} {\bibinfo  {journal} {Physical Review B}\ }\textbf
  {\bibinfo {volume} {102}},\ \bibinfo {pages} {094204} (\bibinfo {year}
  {2020})}\BibitemShut {NoStop}%
\bibitem [{\citenamefont {Cao}\ \emph {et~al.}(2019)\citenamefont {Cao},
  \citenamefont {Tilloy},\ and\ \citenamefont {Luca}}]{Cao2019}%
  \BibitemOpen
  \bibfield  {author} {\bibinfo {author} {\bibfnamefont {X.}~\bibnamefont
  {Cao}}, \bibinfo {author} {\bibfnamefont {A.}~\bibnamefont {Tilloy}},\ and\
  \bibinfo {author} {\bibfnamefont {A.~D.}\ \bibnamefont {Luca}},\ }\bibfield
  {title} {\bibinfo {title} {{Entanglement in a fermion chain under continuous
  monitoring}},\ }\href {https://doi.org/10.21468/SciPostPhys.7.2.024}
  {\bibfield  {journal} {\bibinfo  {journal} {SciPost Phys.}\ }\textbf
  {\bibinfo {volume} {7}},\ \bibinfo {pages} {024} (\bibinfo {year}
  {2019})}\BibitemShut {NoStop}%
\bibitem [{\citenamefont {Chen}\ \emph {et~al.}(2020)\citenamefont {Chen},
  \citenamefont {Li}, \citenamefont {Fisher},\ and\ \citenamefont
  {Lucas}}]{Chen2020emergent}%
  \BibitemOpen
  \bibfield  {author} {\bibinfo {author} {\bibfnamefont {X.}~\bibnamefont
  {Chen}}, \bibinfo {author} {\bibfnamefont {Y.}~\bibnamefont {Li}}, \bibinfo
  {author} {\bibfnamefont {M.~P.~A.}\ \bibnamefont {Fisher}},\ and\ \bibinfo
  {author} {\bibfnamefont {A.}~\bibnamefont {Lucas}},\ }\bibfield  {title}
  {\bibinfo {title} {Emergent conformal symmetry in nonunitary random dynamics
  of free fermions},\ }\href {https://doi.org/10.1103/PhysRevResearch.2.033017}
  {\bibfield  {journal} {\bibinfo  {journal} {Phys. Rev. Res.}\ }\textbf
  {\bibinfo {volume} {2}},\ \bibinfo {pages} {033017} (\bibinfo {year}
  {2020})}\BibitemShut {NoStop}%
\bibitem [{\citenamefont {Buchhold}\ \emph {et~al.}(2021)\citenamefont
  {Buchhold}, \citenamefont {Minoguchi}, \citenamefont {Altland},\ and\
  \citenamefont {Diehl}}]{buchhold2021effective}%
  \BibitemOpen
  \bibfield  {author} {\bibinfo {author} {\bibfnamefont {M.}~\bibnamefont
  {Buchhold}}, \bibinfo {author} {\bibfnamefont {Y.}~\bibnamefont {Minoguchi}},
  \bibinfo {author} {\bibfnamefont {A.}~\bibnamefont {Altland}},\ and\ \bibinfo
  {author} {\bibfnamefont {S.}~\bibnamefont {Diehl}},\ }\bibfield  {title}
  {\bibinfo {title} {Effective theory for the measurement-induced phase
  transition of dirac fermions},\ }\href@noop {} {\bibfield  {journal}
  {\bibinfo  {journal} {Physical Review X}\ }\textbf {\bibinfo {volume} {11}},\
  \bibinfo {pages} {041004} (\bibinfo {year} {2021})}\BibitemShut {NoStop}%
\bibitem [{\citenamefont {Alberton}\ \emph {et~al.}(2021)\citenamefont
  {Alberton}, \citenamefont {Buchhold},\ and\ \citenamefont
  {Diehl}}]{alberton2021entanglement}%
  \BibitemOpen
  \bibfield  {author} {\bibinfo {author} {\bibfnamefont {O.}~\bibnamefont
  {Alberton}}, \bibinfo {author} {\bibfnamefont {M.}~\bibnamefont {Buchhold}},\
  and\ \bibinfo {author} {\bibfnamefont {S.}~\bibnamefont {Diehl}},\ }\bibfield
   {title} {\bibinfo {title} {Entanglement transition in a monitored
  free-fermion chain: From extended criticality to area law},\ }\href@noop {}
  {\bibfield  {journal} {\bibinfo  {journal} {Physical Review Letters}\
  }\textbf {\bibinfo {volume} {126}},\ \bibinfo {pages} {170602} (\bibinfo
  {year} {2021})}\BibitemShut {NoStop}%
\bibitem [{\citenamefont {Hurst}\ \emph {et~al.}(2020)\citenamefont {Hurst},
  \citenamefont {Guo},\ and\ \citenamefont {Spielman}}]{hurst2020feedback}%
  \BibitemOpen
  \bibfield  {author} {\bibinfo {author} {\bibfnamefont {H.~M.}\ \bibnamefont
  {Hurst}}, \bibinfo {author} {\bibfnamefont {S.}~\bibnamefont {Guo}},\ and\
  \bibinfo {author} {\bibfnamefont {I.}~\bibnamefont {Spielman}},\ }\bibfield
  {title} {\bibinfo {title} {Feedback induced magnetic phases in binary
  bose-einstein condensates},\ }\href@noop {} {\bibfield  {journal} {\bibinfo
  {journal} {Physical Review Research}\ }\textbf {\bibinfo {volume} {2}},\
  \bibinfo {pages} {043325} (\bibinfo {year} {2020})}\BibitemShut {NoStop}%
\bibitem [{\citenamefont {Young}\ \emph {et~al.}(2021)\citenamefont {Young},
  \citenamefont {Gorshkov},\ and\ \citenamefont {Spielman}}]{Young2021}%
  \BibitemOpen
  \bibfield  {author} {\bibinfo {author} {\bibfnamefont {J.~T.}\ \bibnamefont
  {Young}}, \bibinfo {author} {\bibfnamefont {A.~V.}\ \bibnamefont
  {Gorshkov}},\ and\ \bibinfo {author} {\bibfnamefont {I.}~\bibnamefont
  {Spielman}},\ }\bibfield  {title} {\bibinfo {title} {Feedback-stabilized
  dynamical steady states in the bose-hubbard model},\ }\href@noop {}
  {\bibfield  {journal} {\bibinfo  {journal} {Physical Review Research}\
  }\textbf {\bibinfo {volume} {3}},\ \bibinfo {pages} {043075} (\bibinfo {year}
  {2021})}\BibitemShut {NoStop}%
\bibitem [{\citenamefont {Yamaguchi}\ \emph {et~al.}(2023)\citenamefont
  {Yamaguchi}, \citenamefont {Hurst},\ and\ \citenamefont
  {Spielman}}]{Yamaguchi2022}%
  \BibitemOpen
  \bibfield  {author} {\bibinfo {author} {\bibfnamefont {E.~P.}\ \bibnamefont
  {Yamaguchi}}, \bibinfo {author} {\bibfnamefont {H.~M.}\ \bibnamefont
  {Hurst}},\ and\ \bibinfo {author} {\bibfnamefont {I.}~\bibnamefont
  {Spielman}},\ }\bibfield  {title} {\bibinfo {title} {Feedback-cooled
  bose-einstein condensation: Near and far from equilibrium},\ }\href@noop {}
  {\bibfield  {journal} {\bibinfo  {journal} {Physical Review A}\ }\textbf
  {\bibinfo {volume} {107}},\ \bibinfo {pages} {063306} (\bibinfo {year}
  {2023})}\BibitemShut {NoStop}%
\bibitem [{\citenamefont {Wu}\ and\ \citenamefont {Eckardt}(2022)}]{Wu2022}%
  \BibitemOpen
  \bibfield  {author} {\bibinfo {author} {\bibfnamefont {L.-N.}\ \bibnamefont
  {Wu}}\ and\ \bibinfo {author} {\bibfnamefont {A.}~\bibnamefont {Eckardt}},\
  }\bibfield  {title} {\bibinfo {title} {Cooling and state preparation in an
  optical lattice via markovian feedback control},\ }\href@noop {} {\bibfield
  {journal} {\bibinfo  {journal} {Physical Review Research}\ }\textbf {\bibinfo
  {volume} {4}},\ \bibinfo {pages} {L022045} (\bibinfo {year}
  {2022})}\BibitemShut {NoStop}%
\bibitem [{\citenamefont {Wampler}\ \emph {et~al.}(2022)\citenamefont
  {Wampler}, \citenamefont {Khor}, \citenamefont {Refael},\ and\ \citenamefont
  {Klich}}]{Wampler2022}%
  \BibitemOpen
  \bibfield  {author} {\bibinfo {author} {\bibfnamefont {M.}~\bibnamefont
  {Wampler}}, \bibinfo {author} {\bibfnamefont {B.~J.}\ \bibnamefont {Khor}},
  \bibinfo {author} {\bibfnamefont {G.}~\bibnamefont {Refael}},\ and\ \bibinfo
  {author} {\bibfnamefont {I.}~\bibnamefont {Klich}},\ }\bibfield  {title}
  {\bibinfo {title} {Stirring by staring: Measurement-induced chirality},\
  }\href@noop {} {\bibfield  {journal} {\bibinfo  {journal} {Physical Review
  X}\ }\textbf {\bibinfo {volume} {12}},\ \bibinfo {pages} {031031} (\bibinfo
  {year} {2022})}\BibitemShut {NoStop}%
\bibitem [{\citenamefont {Mehdi}\ \emph {et~al.}(2023)\citenamefont {Mehdi},
  \citenamefont {Haine}, \citenamefont {Hope},\ and\ \citenamefont
  {Szigeti}}]{mehdi2023fundamental}%
  \BibitemOpen
  \bibfield  {author} {\bibinfo {author} {\bibfnamefont {Z.}~\bibnamefont
  {Mehdi}}, \bibinfo {author} {\bibfnamefont {S.~A.}\ \bibnamefont {Haine}},
  \bibinfo {author} {\bibfnamefont {J.~J.}\ \bibnamefont {Hope}},\ and\
  \bibinfo {author} {\bibfnamefont {S.~S.}\ \bibnamefont {Szigeti}},\
  }\bibfield  {title} {\bibinfo {title} {Fundamental limits of feedback cooling
  ultracold atomic gases},\ }\href@noop {} {\bibfield  {journal} {\bibinfo
  {journal} {arXiv preprint arXiv:2306.09846}\ } (\bibinfo {year}
  {2023})}\BibitemShut {NoStop}%
\bibitem [{\citenamefont {Pezze}\ \emph {et~al.}(2004)\citenamefont {Pezze},
  \citenamefont {Pitaevskii}, \citenamefont {Smerzi}, \citenamefont
  {Stringari}, \citenamefont {Modugno}, \citenamefont {De~Mirandes},
  \citenamefont {Ferlaino}, \citenamefont {Ott}, \citenamefont {Roati},\ and\
  \citenamefont {Inguscio}}]{pezze2004insulating}%
  \BibitemOpen
  \bibfield  {author} {\bibinfo {author} {\bibfnamefont {L.}~\bibnamefont
  {Pezze}}, \bibinfo {author} {\bibfnamefont {L.}~\bibnamefont {Pitaevskii}},
  \bibinfo {author} {\bibfnamefont {A.}~\bibnamefont {Smerzi}}, \bibinfo
  {author} {\bibfnamefont {S.}~\bibnamefont {Stringari}}, \bibinfo {author}
  {\bibfnamefont {G.}~\bibnamefont {Modugno}}, \bibinfo {author} {\bibfnamefont
  {E.}~\bibnamefont {De~Mirandes}}, \bibinfo {author} {\bibfnamefont
  {F.}~\bibnamefont {Ferlaino}}, \bibinfo {author} {\bibfnamefont
  {H.}~\bibnamefont {Ott}}, \bibinfo {author} {\bibfnamefont {G.}~\bibnamefont
  {Roati}},\ and\ \bibinfo {author} {\bibfnamefont {M.}~\bibnamefont
  {Inguscio}},\ }\bibfield  {title} {\bibinfo {title} {Insulating behavior of a
  trapped ideal fermi gas},\ }\href@noop {} {\bibfield  {journal} {\bibinfo
  {journal} {Physical review letters}\ }\textbf {\bibinfo {volume} {93}},\
  \bibinfo {pages} {120401} (\bibinfo {year} {2004})}\BibitemShut {NoStop}%
\bibitem [{\citenamefont {K\"ohl}\ \emph {et~al.}(2005)\citenamefont {K\"ohl},
  \citenamefont {Moritz}, \citenamefont {St\"oferle}, \citenamefont
  {G\"unter},\ and\ \citenamefont {Esslinger}}]{kohl2005fermionic}%
  \BibitemOpen
  \bibfield  {author} {\bibinfo {author} {\bibfnamefont {M.}~\bibnamefont
  {K\"ohl}}, \bibinfo {author} {\bibfnamefont {H.}~\bibnamefont {Moritz}},
  \bibinfo {author} {\bibfnamefont {T.}~\bibnamefont {St\"oferle}}, \bibinfo
  {author} {\bibfnamefont {K.}~\bibnamefont {G\"unter}},\ and\ \bibinfo
  {author} {\bibfnamefont {T.}~\bibnamefont {Esslinger}},\ }\bibfield  {title}
  {\bibinfo {title} {Fermionic atoms in a three dimensional optical lattice:
  Observing fermi surfaces, dynamics, and interactions},\ }\href
  {https://doi.org/10.1103/PhysRevLett.94.080403} {\bibfield  {journal}
  {\bibinfo  {journal} {Phys. Rev. Lett.}\ }\textbf {\bibinfo {volume} {94}},\
  \bibinfo {pages} {080403} (\bibinfo {year} {2005})}\BibitemShut {NoStop}%
\bibitem [{\citenamefont {Moritz}\ \emph {et~al.}(2005)\citenamefont {Moritz},
  \citenamefont {St\"oferle}, \citenamefont {G\"unter}, \citenamefont
  {K\"ohl},\ and\ \citenamefont {Esslinger}}]{moritz2005confinement}%
  \BibitemOpen
  \bibfield  {author} {\bibinfo {author} {\bibfnamefont {H.}~\bibnamefont
  {Moritz}}, \bibinfo {author} {\bibfnamefont {T.}~\bibnamefont {St\"oferle}},
  \bibinfo {author} {\bibfnamefont {K.}~\bibnamefont {G\"unter}}, \bibinfo
  {author} {\bibfnamefont {M.}~\bibnamefont {K\"ohl}},\ and\ \bibinfo {author}
  {\bibfnamefont {T.}~\bibnamefont {Esslinger}},\ }\bibfield  {title} {\bibinfo
  {title} {Confinement induced molecules in a 1d fermi gas},\ }\href
  {https://doi.org/10.1103/PhysRevLett.94.210401} {\bibfield  {journal}
  {\bibinfo  {journal} {Phys. Rev. Lett.}\ }\textbf {\bibinfo {volume} {94}},\
  \bibinfo {pages} {210401} (\bibinfo {year} {2005})}\BibitemShut {NoStop}%
\bibitem [{\citenamefont {Giorgini}\ \emph {et~al.}(2008)\citenamefont
  {Giorgini}, \citenamefont {Pitaevskii},\ and\ \citenamefont
  {Stringari}}]{giorgini2008theory}%
  \BibitemOpen
  \bibfield  {author} {\bibinfo {author} {\bibfnamefont {S.}~\bibnamefont
  {Giorgini}}, \bibinfo {author} {\bibfnamefont {L.~P.}\ \bibnamefont
  {Pitaevskii}},\ and\ \bibinfo {author} {\bibfnamefont {S.}~\bibnamefont
  {Stringari}},\ }\bibfield  {title} {\bibinfo {title} {Theory of ultracold
  atomic fermi gases},\ }\href@noop {} {\bibfield  {journal} {\bibinfo
  {journal} {Reviews of Modern Physics}\ }\textbf {\bibinfo {volume} {80}},\
  \bibinfo {pages} {1215} (\bibinfo {year} {2008})}\BibitemShut {NoStop}%
\bibitem [{\citenamefont {Hurst}\ and\ \citenamefont
  {Spielman}(2019)}]{hurst2019measurement}%
  \BibitemOpen
  \bibfield  {author} {\bibinfo {author} {\bibfnamefont {H.~M.}\ \bibnamefont
  {Hurst}}\ and\ \bibinfo {author} {\bibfnamefont {I.}~\bibnamefont
  {Spielman}},\ }\bibfield  {title} {\bibinfo {title} {Measurement-induced
  dynamics and stabilization of spinor-condensate domain walls},\ }\href@noop
  {} {\bibfield  {journal} {\bibinfo  {journal} {Physical Review A}\ }\textbf
  {\bibinfo {volume} {99}},\ \bibinfo {pages} {053612} (\bibinfo {year}
  {2019})}\BibitemShut {NoStop}%
\bibitem [{\citenamefont {Kraus}(1974)}]{Kraus1974}%
  \BibitemOpen
  \bibfield  {author} {\bibinfo {author} {\bibfnamefont {K.}~\bibnamefont
  {Kraus}},\ }\bibfield  {title} {\bibinfo {title} {Operations and effects in
  the hilbert space formulation of quantum theory},\ }in\ \href@noop {} {\emph
  {\bibinfo {booktitle} {Foundations of quantum mechanics and ordered linear
  spaces}}}\ (\bibinfo  {publisher} {Springer},\ \bibinfo {year} {1974})\ pp.\
  \bibinfo {pages} {206--229}\BibitemShut {NoStop}%
\bibitem [{\citenamefont {Jacobs}\ and\ \citenamefont
  {Steck}(2006)}]{Jacobs2006}%
  \BibitemOpen
  \bibfield  {author} {\bibinfo {author} {\bibfnamefont {K.}~\bibnamefont
  {Jacobs}}\ and\ \bibinfo {author} {\bibfnamefont {D.~A.}\ \bibnamefont
  {Steck}},\ }\bibfield  {title} {\bibinfo {title} {A straightforward
  introduction to continuous quantum measurement},\ }\href@noop {} {\bibfield
  {journal} {\bibinfo  {journal} {Contemporary Physics}\ }\textbf {\bibinfo
  {volume} {47}},\ \bibinfo {pages} {279} (\bibinfo {year} {2006})}\BibitemShut
  {NoStop}%
\bibitem [{Note1()}]{Note1}%
  \BibitemOpen
  \bibinfo {note} {This discrete update model can be connected to a continuous
  stochastic equation as follows. To consider a measurement of duration $dt$,
  we redefine our measurement strength according to $\varphi \rightarrow
  \varphi \protect \sqrt {dt}$. This allows us to introduce Weiner increment
  $dW = \protect \sqrt {dt}\Delta m$ with the correct statistical properties of
  $\protect \overline {dW} = 0$ and $\protect \overline {dWdW'} = \protect
  \overline {\Delta m \Delta m'} dt = dt$.}\BibitemShut {Stop}%
\bibitem [{\citenamefont {Brezini}\ and\ \citenamefont
  {Zekri}(1992)}]{Brezini1992}%
  \BibitemOpen
  \bibfield  {author} {\bibinfo {author} {\bibfnamefont {A.}~\bibnamefont
  {Brezini}}\ and\ \bibinfo {author} {\bibfnamefont {N.}~\bibnamefont
  {Zekri}},\ }\bibfield  {title} {\bibinfo {title} {Overview on some aspects of
  the theory of localization},\ }\href@noop {} {\bibfield  {journal} {\bibinfo
  {journal} {phys. stat. sol. (b)}\ }\textbf {\bibinfo {volume} {169}},\
  \bibinfo {pages} {253} (\bibinfo {year} {1992})}\BibitemShut {NoStop}%
\bibitem [{\citenamefont {Evers}\ and\ \citenamefont
  {Mirlin}(2008)}]{Evers2008}%
  \BibitemOpen
  \bibfield  {author} {\bibinfo {author} {\bibfnamefont {F.}~\bibnamefont
  {Evers}}\ and\ \bibinfo {author} {\bibfnamefont {A.~D.}\ \bibnamefont
  {Mirlin}},\ }\bibfield  {title} {\bibinfo {title} {Anderson transitions},\
  }\href@noop {} {\bibfield  {journal} {\bibinfo  {journal} {Reviews of Modern
  Physics}\ }\textbf {\bibinfo {volume} {80}},\ \bibinfo {pages} {1355}
  (\bibinfo {year} {2008})}\BibitemShut {NoStop}%
\bibitem [{\citenamefont {Sansonetti}\ \emph {et~al.}(2011)\citenamefont
  {Sansonetti}, \citenamefont {Simien}, \citenamefont {Gillaspy}, \citenamefont
  {Tan}, \citenamefont {Brewer}, \citenamefont {Brown}, \citenamefont {Wu},\
  and\ \citenamefont {Porto}}]{Sansonetti2011}%
  \BibitemOpen
  \bibfield  {author} {\bibinfo {author} {\bibfnamefont {C.~J.}\ \bibnamefont
  {Sansonetti}}, \bibinfo {author} {\bibfnamefont {C.~E.}\ \bibnamefont
  {Simien}}, \bibinfo {author} {\bibfnamefont {J.~D.}\ \bibnamefont
  {Gillaspy}}, \bibinfo {author} {\bibfnamefont {J.~N.}\ \bibnamefont {Tan}},
  \bibinfo {author} {\bibfnamefont {S.~M.}\ \bibnamefont {Brewer}}, \bibinfo
  {author} {\bibfnamefont {R.~C.}\ \bibnamefont {Brown}}, \bibinfo {author}
  {\bibfnamefont {S.}~\bibnamefont {Wu}},\ and\ \bibinfo {author}
  {\bibfnamefont {J.~V.}\ \bibnamefont {Porto}},\ }\bibfield  {title} {\bibinfo
  {title} {Absolute transition frequencies and quantum interference in a
  frequency comb based measurement of the $^{6,7}\mathrm{Li}$ $d$ lines},\
  }\href@noop {} {\bibfield  {journal} {\bibinfo  {journal} {Phys. Rev. Lett.}\
  }\textbf {\bibinfo {volume} {107}},\ \bibinfo {pages} {023001} (\bibinfo
  {year} {2011})}\BibitemShut {NoStop}%
\bibitem [{Note2()}]{Note2}%
  \BibitemOpen
  \bibinfo {note} {Because the Fermi-sea ground state has spatial correlations
  on the scale of $1/k_{\protect \rm F}$ it does not uniformly sample Fock
  space, and the associated IPR is slightly larger than zero.}\BibitemShut
  {Stop}%
\bibitem [{Note3()}]{Note3}%
  \BibitemOpen
  \bibinfo {note} {Let $\protect \{\protect \mathcal D_i\protect \}_i$ be the
  set of subspaces, we define $\protect \overline {D}_{\protect \rm s} \equiv
  \protect \overline {{\protect \rm Dim}(\protect \mathcal D_i)}$.}\BibitemShut
  {Stop}%
\bibitem [{\citenamefont {Sala}\ \emph {et~al.}(2020)\citenamefont {Sala},
  \citenamefont {Rakovszky}, \citenamefont {Verresen}, \citenamefont {Knap},\
  and\ \citenamefont {Pollmann}}]{sala2020ergodicity}%
  \BibitemOpen
  \bibfield  {author} {\bibinfo {author} {\bibfnamefont {P.}~\bibnamefont
  {Sala}}, \bibinfo {author} {\bibfnamefont {T.}~\bibnamefont {Rakovszky}},
  \bibinfo {author} {\bibfnamefont {R.}~\bibnamefont {Verresen}}, \bibinfo
  {author} {\bibfnamefont {M.}~\bibnamefont {Knap}},\ and\ \bibinfo {author}
  {\bibfnamefont {F.}~\bibnamefont {Pollmann}},\ }\bibfield  {title} {\bibinfo
  {title} {Ergodicity breaking arising from hilbert space fragmentation in
  dipole-conserving hamiltonians},\ }\href@noop {} {\bibfield  {journal}
  {\bibinfo  {journal} {Physical Review X}\ }\textbf {\bibinfo {volume} {10}},\
  \bibinfo {pages} {011047} (\bibinfo {year} {2020})}\BibitemShut {NoStop}%
\bibitem [{\citenamefont {Moudgalya}\ \emph {et~al.}(2022)\citenamefont
  {Moudgalya}, \citenamefont {Bernevig},\ and\ \citenamefont
  {Regnault}}]{moudgalya2022quantum}%
  \BibitemOpen
  \bibfield  {author} {\bibinfo {author} {\bibfnamefont {S.}~\bibnamefont
  {Moudgalya}}, \bibinfo {author} {\bibfnamefont {B.~A.}\ \bibnamefont
  {Bernevig}},\ and\ \bibinfo {author} {\bibfnamefont {N.}~\bibnamefont
  {Regnault}},\ }\bibfield  {title} {\bibinfo {title} {Quantum many-body scars
  and hilbert space fragmentation: a review of exact results},\ }\href@noop {}
  {\bibfield  {journal} {\bibinfo  {journal} {Reports on Progress in Physics}\
  }\textbf {\bibinfo {volume} {85}},\ \bibinfo {pages} {086501} (\bibinfo
  {year} {2022})}\BibitemShut {NoStop}%
\end{thebibliography}%

\end{document}